\documentclass[preprintnumber, amsmath, amssymb, aps, floatfix]{revtex4-2}

\usepackage{graphicx}% Include figure files
\usepackage{dcolumn}% Align table columns on decimal point
\usepackage{bm}% bold math
\usepackage{dcolumn} %align tables at decimal point
\usepackage{longtable} %handle multipage tables
\usepackage{makecell} %properly format table headings with text wrap
\usepackage{multirow}
\usepackage[english]{babel}
\usepackage{lineno} %add linenumbers
\usepackage{booktabs} %help with table formatting
\usepackage{tabularx}
\newcommand{\figref}[1]{Fig.~\ref{#1}}
\newcommand{\tabref}[1]{Table~\ref{#1}}

\renewcommand{\eqref}[1]{Eq.~(\ref{#1})}
\makeatletter
\def\bbl@set@language#1{%
  \edef\languagename{%
    \ifnum\escapechar=\expandafter`\string#1\@empty
    \else\string#1\@empty\fi}%
  \@ifundefined{babel@language@alias@\languagename}{}{%
    \edef\languagename{\@nameuse{babel@language@alias@\languagename}}%
  }%
  \select@language{\languagename}%
  \expandafter\ifx\csname date\languagename\endcsname\relax\else
    \if@filesw
      \protected@write\@auxout{}{\string\select@language{\languagename}}%
      \bbl@for\bbl@tempa\BabelContentsFiles{%
        \addtocontents{\bbl@tempa}{\xstring\select@language{\languagename}}}%
      \bbl@usehooks{write}{}%
    \fi
  \fi}
\newcommand{\DeclareLanguageAlias}[2]{%
  \global\@namedef{babel@language@alias@#1}{#2}%
}
\makeatother

\DeclareLanguageAlias{en}{english}
\DeclareLanguageAlias{eng}{english}
\DeclareLanguageAlias{en-CA}{english}

\begin{document}

\title{Beyond gender: The intersectional impact of community demographics on the continuation rates of male and female students into high school physics}
%\thanks{A footnote to the article title}

\author{Eamonn Corrigan}
% \email{eamonn@uoguelph.ca}
\author{Martin Williams}
 \email{martin.williams@uoguelph.ca}
\affiliation{
 Department of Physics, University of Guelph, Guelph, Ontario, N1G 2W1, Canada
 }
 \author{Mary A. Wells}
\affiliation{
 Faculty of Engineering, University of Waterloo, Waterloo, Ontario, N2L 3G1, Canada
 }
 
\date{\today}% It is always \today, today,

\begin{abstract}
This study examines the complex interplay of gender and other demographics on continuation rates in high school physics. Using a diverse dataset that combines demographics from the Canadian Census and eleven years of gendered enrolment data from the Ontario Ministry of Education, we track student cohorts as they transition from mandatory science to elective physics courses. We then employ hierarchical linear modelling to quantify the interaction effects between gender and other demographics, providing a detailed perspective on the on continuation in physics. Our results indicate the racial demographics of a school’s neighbourhood have a limited impact on continuation once controlling for other factors such as socioeconomic status, though neighbourhoods with a higher Black population were a notable exception, consistently exhibiting significantly lower continuation rates for both male and female students. A potential role model effect related to parental education was also found as the proportion of parents with STEM degrees correlates positively with increased continuation rates, whereas an increase in non-STEM degrees corresponds with a reduced SCR. The most pronounced effects are school-level factors. Continuation rates in physics are very strongly correlated with continuation in chemistry or calculus – effects which are much stronger for male than female students. Conversely, continuation in biology positively correlates with the continuation of female students in physics, with little to no effect found for male students. Nevertheless, the effect sizes observed for chemistry and calculus markedly outweigh that for biology. This is further evidence that considering STEM as a homogeneous subject when examining gender disparities is misguided. These insights can guide future education policies and initiatives to increase continuation rates and foster greater gender equity and inclusivity in physics education. 
\end{abstract}


\maketitle

%\tableofcontents

\newpage

\section{\label{sec:census-intro} Introduction}

\begin{figure}[hb!]
\includegraphics[width = 0.6\linewidth]{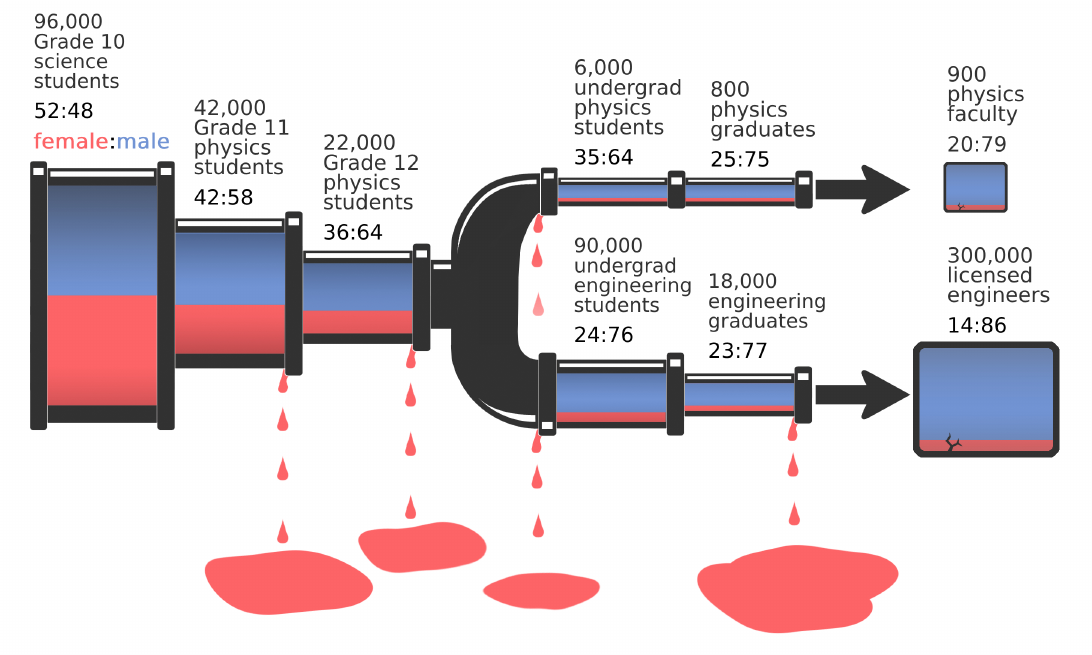}
\caption[Leaky pipeline into physics and engineering]{Ontario's leaky pipeline of women in physics and engineering. The pipeline progresses left to right from grade 10 science, the last mandatory high school course, to undergraduate education and beyond. The blue represents males while red is for females; top and bottom of the pipe respectively. Data is compiled from the Ontario Ministry of Education, from the Canadian Association of Physicists, and Engineers Canada. Data for high school and engineering are from 2018 for consistency, while the physics data is from 2022, the only year of available data.}
\label{fig:Pipeline}
\end{figure}

Women continue to be underrepresented in STEM (science, technology, engineering, and mathematics) majors and careers, with this underrepresentation varying significantly across different fields. While biology has seen a positive shift, with women achieving equal or greater representation, significant gender gaps persist in fields like physics and engineering \cite{aps_bachelors_2022, huang_historical_2020}. These gaps are evident from secondary school on-wards, impacting students' intentions in STEM fields at the undergraduate level and beyond  (see \figref{fig:Pipeline}) \cite{krakehl_intersectional_2021, hazari_connecting_2010, card_high_2021, corrigan_high_2023}. For example, in Ontario's grade 12 physics classes, female participation has remained consistently low, whereas in biology, female representation has increased substantially \cite{corrigan_high_2023}. Efforts to address the gender gap in engineering and physics in university programs and in the workforce will remain limited by this gap in high school participation and thus it is paramount to understand fully all contributing factors.

Recent studies in physics education have started to incorporate intersectionality, examining how gender interacts with other demographic factors \cite{krakehl_intersectional_2021, hennessey_workshop_2019, becares_understanding_2015}. Using eleven years of gendered administrative data from the Ontario Ministry of Education data combined with community-level demographics from the Canadian Census, we built predictive models about the influence of community demographics on the continuation rates of male and female students in high school physics. 

\subsection{\label{sec:lit_review}Literature Review: Community-Level Demographics}

Racialized and Indigenous individuals face significant disparities in STEM education and career outcomes. These disparities manifest in various forms, including lower levels of degree attainment in physics and engineering \cite{american_physical_society_physics_2020, smolina_can_2021, espinosa_pipelines_2009, ma_race_2015}, and worse workforce outcomes both in terms of hiring and participation as well as advancement and promotions \cite{polyzou_indigenous_2022, eaton_how_2019, arriagada_achievements_2021}. Many of these disparities are most prominently found for Indigenous and Black communities, with outcomes worse for women compared with men \cite{polyzou_indigenous_2022, arriagada_achievements_2021, espinosa_pipelines_2009}

Students’ socio-economic status (SES) also significantly impacts educational outcomes. Children are more likely to attend university if they come from higher-income households \cite{walpole_socioeconomic_2003, frenette_postsecondary_2017}, though higher SES students often select majors based on cultural norms or preferences, while lower SES students tend to prioritize economic factors \cite{brand_who_2010}. As STEM careers tend to be higher paid, this aligns with research showing that lower SES students who did enrol in post-secondary education were more confident in their choice of STEM as a major compared with their higher SES peers \cite{lichtenberger_predicting_2012}. 

Independent of income, children of university-educated parents are more likely to pursue and complete higher education \cite{cataldi_first-generation_2018}, potentially hinting at a role model effect or ingrained cultural expectations. This is supported by research which showed that parents’ gendered expectations about their child’s occupation at the age of 15 are strongly correlated to the actual career of the child at 28 \cite{jacobs_enduring_2006}, while parental views about physics and its usefulness for employment predict a child’s likelihood to enrol in senior level physics at high school \cite{jones_examining_2022}. Furthermore, neighbourhoods with a higher proportion of women working in STEM had greater female participation in advanced-level high school STEM courses \cite{riegle-crumb_gender_2014}. Clearly, parents and those in a student's community can significantly influence their educational choices.

\subsection{\label{sec:intent}The Intent of this Work}
The research above describes the complicated picture that is STEM education. Variables such as SES, familial education, and racial identity all significantly shape educational outcomes - and each of these elements is further complicated by their intersections with gender. In this research, we aim to explore how the demographic composition of a school community and how the intersectional dynamics of these community demographics and gender may affect student outcomes. we have formalized these ideas into two distinct research questions (RQs):

\begin{quote}
    \textbf{RQ1}: Do community-level demographic factors significantly predict Student Continuation Rates into grade 11 and 12 physics?

    \textbf{RQ2}: Which community-level demographic factors predict changes in the average gender gap in Student Continuation Rates between male and female students?
\end{quote}

In this work, we use administrative data obtained from the Ontario Ministry of Education combined with community-level demographics from the Canadian Census to try and answer both research questions.

\section{\label{Methods} Materials \& Methods}
\subsection{\label{subsec:enrol_data} Student Enrolment Dataset}

Our analysis was focused on the province of Ontario, making up almost $40\%$ of Canada’s population \cite{government_of_ontario_census_2022-2}. Ontario is a diverse province that reflects the demographic makeup of Canada as a whole, including a range of racial and ethnic groups, urban and rural populations, and a diverse range of immigration statuses and socioeconomic backgrounds \cite{government_of_ontario_census_2022, government_of_ontario_census_2022-1, government_of_ontario_census_2022-2, government_of_ontario_census_2022-3}. This diversity allows for a representative sample of the Canadian population and suggests our findings may be generalized to a wider population in similar Western nations. 

The data set used in this analysis was obtained through a research partnership with the Ontario Ministry of Education (OME). It contains the total number of male and female students who enrolled in each university stream science and math course from grade 10 to grade 12 for every secondary school across the province ($N \approx 840$, but varies year to year as schools open or close). In this context, university stream means these courses are accepted for admission to Canadian universities. The dataset includes all grade 12 science and math courses commonly required as prerequisites to undergraduate STEM degrees, as well as the grade 10 and grade 11 prerequisites. The exact courses included are shown in \tabref{tab:courses}. The data spans 11 years, from the 2007/08 academic year through 2017/18. 

This research project was reviewed and approved by the University of Guelph Research Ethics Board (REB \#19-06-015), receiving an exemption from a full ethics review. This exemption was based on the aggregate nature of the data requested and our provision for the ministry to suppress low enrolment counts to further protect student privacy. In accordance with this exemption, the Ontario Ministry of Education (OME) chose to suppress all enrolment counts with $<10$ students. For this work, suppressed cells have been removed.

\subsection{\label{subsec:SCR} Tracking Student Cohorts: Student Continuation Rate}
The progress of student cohorts moving through high school was tracked by calculating the rate at which students stayed in STEM classes (Student Continuation Rate; SCR) after their last mandatory STEM credits. SCR is calculated as the ratio of male/female enrolment in grade 11 physics divided by total enrolment in grade 10 science from one year prior; grade 12 enrolments are divided by grade 10 science enrolment from two years prior. SCR was calculated for all schools in Ontario for all years available in our dataset. Male and female SCRs were calculated separately.

Tracking cohorts like this is not a perfect representation – some students will move between schools and others may not take their courses in a linear fashion year to year – but this still provides a good estimate of the continuation rate for most students. For some schools, year-to-year data was missing or suppressed and so SCR could not be calculated for all schools and all years. With these removed, our finale sample had $N_m=5802$ and $N_f=5360$ observations for male and female students in grade 11 physics and $N_m=4526; N_f=3027$ for grade 12 physics. These figures represent the total number of SCR data entries from individual schools across all years of data. In both instances, the sample size for male students was larger than that for female students, whose data were more likely to be suppressed due to low enrolment. There were also a few outliers where SCR $> 100\%$. Upon closer examination, we found these were primarily caused by school restructurings or closures where student cohorts were combined. Representing only $<0.5\%$ of the total sample, these outliers were removed from the analysis.

\begin{table}[ht]
\centering
\caption{Ontario high school STEM courses for which enrolment data was obtained from 2007/2008 through to 2017/2018}
\label{tab:courses}
\begin{tabular}{ll}
\toprule
\textbf{Science Courses} & \textbf{Math Courses} \\
\midrule
Gr 10 Academic Science (SNC2D)* & Gr 10 Academic Principles of Mathematics (MPM2D) \\
Gr 11 University Biology (SBI3U) & Gr 11 University Functions (MCR3U)* \\
Gr 11 University Chemistry (SCH3U) & Gr 12 University Advanced Functions (MHF4U)$^{\dagger}$ \\
Gr 11 University Physics (SPH3U) & Gr 12 University Calculus and Vectors (MCV4U)$^{\dagger}$ \\
Gr 12 University Biology (SBI4U)$^{\dagger}$ & \\
Gr 12 University Chemistry (SCH4U)$^{\dagger}$ & \\
Gr 12 University Physics (SPH4U)$^{\dagger}$ & \\
\midrule
\multicolumn{2}{p{\dimexpr \linewidth-2\tabcolsep}}{* These courses satisfy mandatory requirements for high school graduation.} \\
\multicolumn{2}{p{\dimexpr \linewidth-2\tabcolsep}}{$^{\dagger}$ These courses are required for entry into most undergraduate physics and engineering programs across Canada.} \\
\end{tabular}
\end{table}

\subsection{\label{subsec:cenus_data} Census Demographic Data}

\begin{figure}
\includegraphics[width = \linewidth]{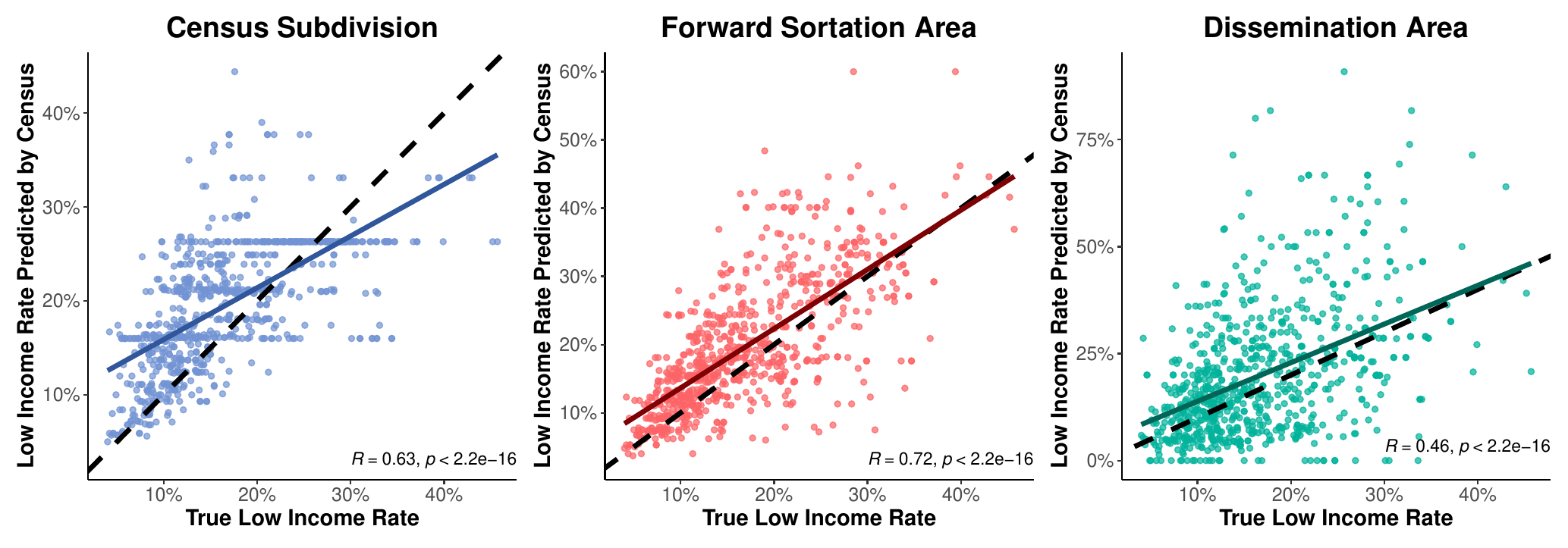}
\caption[Goodness of fit for different Census Geographic Units]{Scatter plots showing the true proportion of students coming from low-income households as reported by the Ontario Ministry of Education versus the proportion predicted by linking the Canadian census using three different geographic units: Census Subdivision, Forward Sortation Area, and Dissemination Area. The dotted black line on each graph indicates a perfect match between the true and predicted values. The closer each point is to the dotted black line, the smaller the error between the predicted and actual values. The solid coloured lines come from a linear regression fit. Pearson correlation for each of the three geographic units is also reported.}
\label{fig:Census_Areas}
\end{figure}

School selection is primarily dictated by geographic location, but the catchment areas used by each school do not map directly to the geographic units used by the Canadian Census. Thus, we sought to determine which census geographic unit best correlated with the school’s true population. The data set provided by the OME included the postal code for each high school which was used to link census data using the Postal Code Conversion File Plus (PCCF+) v7.2 \cite{statistics_canada_postal_2017}. The OME also annually releases a dataset of “School Information and Student Demographics” that includes information about each school and its population \cite{ontario_ministry_of_education_school_2018}. Importantly, this included the proportion of low-income students as defined by the Low-Income Measure, After Tax (LIM-AT), which is also reported in the Canadian Census. This provided us with a variable to measure the correlation between the community demographics as predicted by different census geographic areas and the true demographics of each school.

Three different census geographic areas – Forward Sortation Area (FSA), Census Subdivision, and Census Dissemination Area – were considered. Pearson’s Correlation Coefficient was calculated comparing LIM-AT for each school as reported by the OME and LIM-AT as predicted by each census geographic area \figref{fig:Census_Areas}.  FSA was the most highly correlated  $(r = 0.72;p<2.2 \times 10^{-16})$ and was chosen as the best proxy for each school’s catchment area. FSA has also been used in similar research to model student demographics and the effect on STEM education choices within the Ontario high school to post-secondary transition \cite{dooley_understanding_2017}. Finally, we used data from both the 2011 and 2016 censuses. School enrolment data was linked to which ever census was closest in time.

\subsection{\label{subsec:vars} Variable Selection}

\subsubsection{\label{sssec:race_vars} Race and Indigeneity}

\begin{figure}
\includegraphics[width = 0.8\linewidth]{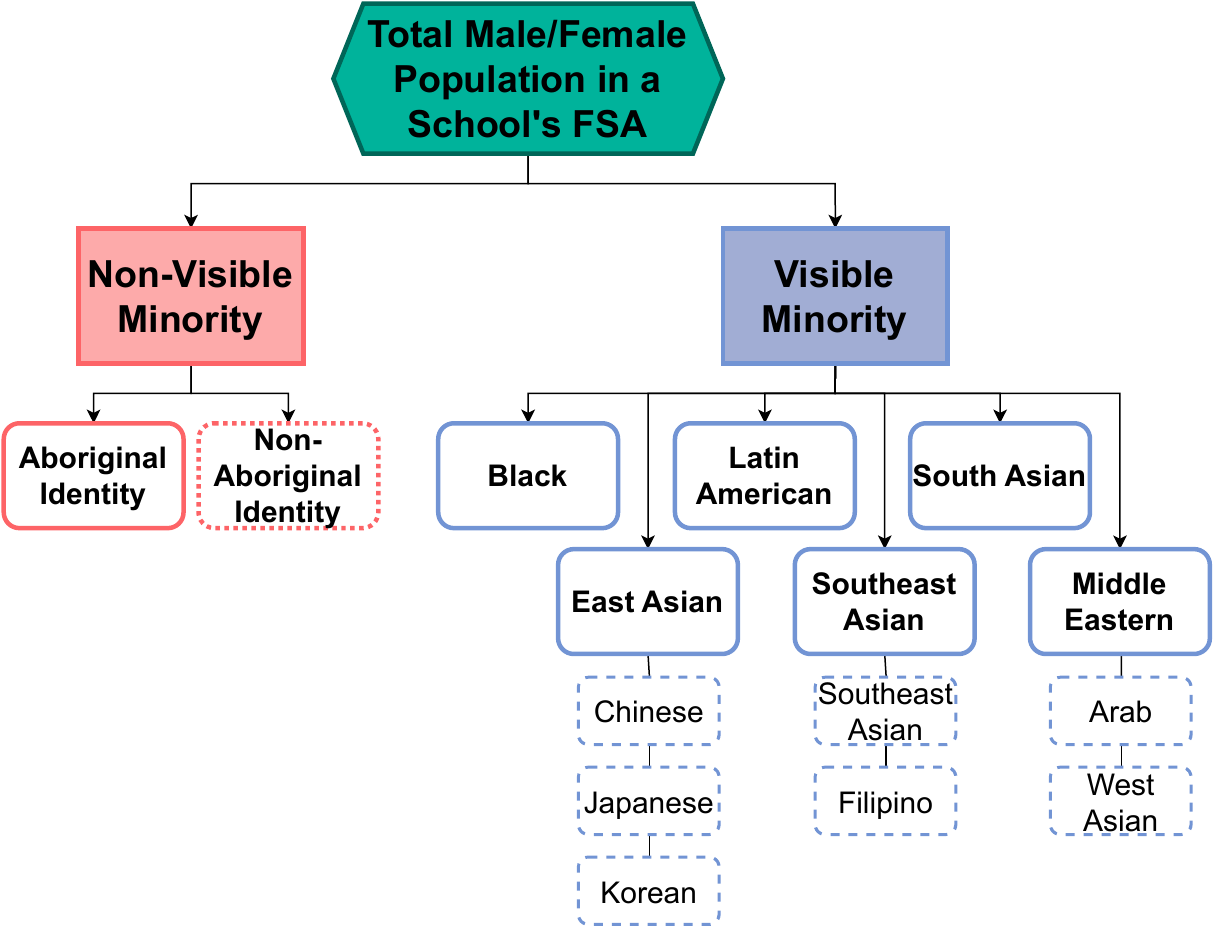}
\caption[Groupings of race-based and Indigenous identities for the Canadian Census]{Tree diagram showing the grouping structure of the Canadian Census variables used to model the ethnic/racial demographics of a given school. The total male and female populations of the FSA surrounding each school (filled, green hexagon) is split into those who do and do not identify as visible minorities (filled blue and red rectangles; right and left branches respectively). Subdividing each of these two categories, we have the 7 variables which are used as predictor variables in our model (rounded rectangles with solid borders) as well as our reference category (rounded, red rectangle with dotted border). The Visible Minority subgroups used in our regression analysis are shown with the Census Population Groups that make them up (dashed, blue, rounded rectangles; where applicable).}
\label{fig:Pop_group_vars}
\end{figure}

The Canadian Census reports two broad categories related to racial demographics: \textit{Visible Minorities}\footnote{The term Visible Minority was defined by the Employment Equity Act as “persons, other than Aboriginal peoples, who are non-Caucasian in race or non-white in colour”. Beyond this section where we use variable names as given by the census, we chose to use the more widely accepted terms Person of Colour or Racialized Individual instead of Visible Minority and Indigenous instead of Aboriginal.} and \textit{Non-Visible Minorities}. The Canadian census further subdivides \textit{Visible Minorities} into 12 different population groups. Two of these, \textit{Multiple Visible Minorities} (the respondent selected more than one option) and \textit{Not Included Elsewhere} (a write-in option) were both removed. The remaining 10 population groups were further simplified to only six groups for parsimony and to increase statistical power. Groupings were decided based on multicollinearity among groups and geographic proximity of original descent. After the fact, these combined groupings were found to align with those recommended by the Canadian Institute for Health Information \cite{canadian_institute_for_health_information_proposed_2020}. Those who identify as Not a Visible Minority are also subdivided by the census as those with an Aboriginal Identity and those with Non-Aboriginal Identity. This results in six subgroups of \textit{Visible Minority} plus \textit{Aboriginal} which make up the seven predictor variables used in our model to examine the influence of race or Indigeneity on continuation rates. The exact grouping used are shown in Figure \ref{fig:Pop_group_vars}.

\subsubsection{\label{sssec:SES_Vars} Socioeconomic Factors}

The census variables selected included the proportion of individuals under 18 living in low-income households as measured by LIM-AT as well as the logarithm of median family income of single parents and couples with children. The choice to include both income variables was made as we wanted to separately examine the possible effects of living in a low-income household from the known effects of higher income families being more likely to attend university \cite{walpole_socioeconomic_2003, frenette_postsecondary_2017}.

We subsequently included the proportion of respondents aged 25-64 working in a STEM field, and the proportion of respondents over 15 who have obtained a post-secondary degree in a STEM-related field. Here, we chose to include both variables as a large proportion of people who study STEM, do not end up working in these fields. Prior work has found that having an increased proportion of women working in STEM boost high school physics enrolment for female students \cite{riegle-crumb_gender_2014}. However, most people who obtain a physics degree do not end up working in that field \cite{aip_statistics_initial_2022}. Thus, we hypothesized that the proportion of individuals working in STEM and having been educated in STEM could separately influence student continuation. Finally, we also include the proportion of respondents over 15 who have obtained a post-secondary degree in any field to control for the known effect that children of parents who have attended university are more likely to do the same
\cite{cataldi_first-generation_2018, turcotte_intergenerational_2011}.

The usage of “STEM-related field” for occupation or education above was defined as follows. The National Occupational Classifications category “Natural and Applied Sciences and Related Occupations” from the census was used for employment \cite{statistics_canada_national_2016}. Post-secondary degrees were sorted using the Classification of Instructional Programs from the Canadian Census \cite{statistics_canada_classification_2016}. All respondents in Category 06 – Physical and Life Sciences and Technologies, Category 07 – Mathematics, Computer and Information Sciences, and Category 08 - Architecture, Engineering, and Related Technologies were grouped as having a degree in STEM for analysis. We excluded Category 10 – Health and Related Fields as our primary areas of interest are physics, engineering, and related fields.

\subsubsection{Addressing Potential Multicollinearity}
When considering multiple related variables, as we have chosen to do, there is always a risk of multicollinearity impacting the statistical results. We checked the levels of multicollinearity observed in our combined models by calculating the Variance Inflation Factor and all values were found to be <5.0, indicating multicollinearity is unlikely to negatively impact our modelling \cite{stine_graphical_1995}.

\subsubsection{Standardizing Variables}

As many of the predictor variables are measured on different scales, all census variables were standardized to have a mean of 0 and a standard deviation of 1. Hence, when used in linear modelling, the associated regression estimates will predict the expected change in physics SCR for a one standard deviation change in each predictor variable. This allows for the direct comparison of effect sizes between variables measured on different scales.

\subsection{\label{subsec:models} Model Selection}

To address RQ1, we used multiple linear regression to measure how various community-level demographic variables, as measured by the census, predict SCR into grade 11 and grade 12 physics. For these two models, male and female students are grouped together, focusing only on the first level effects of  community-level demographics on continuation.

The OME runs both public and Catholic secondary schools where students at the latter are required to take religion courses for graduation, limiting elective options for students. To account for this, we utilized mixed effects linear regression to model SCR, introducing a random effect variable included to control for systematic differences between the public and Catholic school systems.  Additionally, a random effect term for year was added, as previous research has shown that SCR has been measurably changing over time \cite{corrigan_high_2023}. We opted to treat variations over time as a random effect rather than a fixed effect, as this resulted in better model fits as measured by a lower Akaike Information Criterion (AIC) values. Additionally, this allowed us to adjust for the variability of each individual year rather than fitting all years to a single slope parameter. The resulting model to address RQ1 takes the form:

\begin{equation} \label{eq:census_model}
    Y_{ij} = \beta_0 + \beta_1 X_1 + \beta_2 X_2 + \ldots + \beta_n X_n + S_i + T_j + \epsilon.
\end{equation}

Here, $Y_{ij}$ is our predicted outcome variable, Student Continuation Rate in grade 11 or grade 12 physics. The fixed effect predictor variables, i.e., the community-level demographic factors, are represented by $\{X_1, X_2, \ldots, X_n\}$, while $\{\beta_1, \beta_2, \ldots, \beta_n\}$ represent the corresponding regression estimates. $S_i$ denotes the random effect accounting for the average change in SCR between Public and Catholic schools in Ontario, while $T_j$ represents the random effect for Year $j$, controlling for variations in SCR year to year. Finally, $\beta_0$ is the estimated global intercept.

For RQ2, we first calculate the difference in continuation rates  between male and female students in grade 11 and 12 physics at each school. Positive values indicate SCR is higher for males, while a negatives indicate higher continuation for female students. Once again, we employ the mixed-effects regression model of \eqref{eq:census_model} to account for differences between school systems and changes over time, but the outcome variable, $Y_{ij}$, is now the gender gap in SCR between male and female students. A negative regression estimate thus indicates the gender gap in SCR into physics is smaller than average.

In summary, we created four mixed-effects linear regression models. The first two, which take the form of \eqref{eq:census_model}, predict SCRs into grade 11 and grade 12 physics for all students. This directly addresses RQ1. The second two models also take the form of \eqref{eq:census_model}, but the outcome variable, $Y_{ij}$, is now the difference in SCR between male and female students into grade 11 and 12 physics. In all models the predictor variables are community-level demographic variables from the census, all census variables have been standardized, and random effects are included to control for differences between the public and Catholic school systems as well as year of observation.

All analyses were performed using R Statistical Software version 4.3.1 \cite{r_core_team_r_2023} in conjunction with the Tidyverse \cite{wickham_welcome_2019}, lme4 \cite{bates_fitting_2015}, Cowplot \cite{wilke_cowplot_2020}, and Stargazer \cite{hlavac_stargazer_2022} packages. Finally, as an extension for lme4, conditional $R^2$ values were calculated for our models using the MuMin package \cite{barton_mumin_2023}, following the statistical procedure described in \cite{nakagawa_coefficient_2017}.

\section{Results}\label{sec:census_results}

\subsection{RQ1 - Community-Level Demographics Predict SCR}\label{subsec:RQ1}
\begin{figure}
\includegraphics[width = 0.85\linewidth]{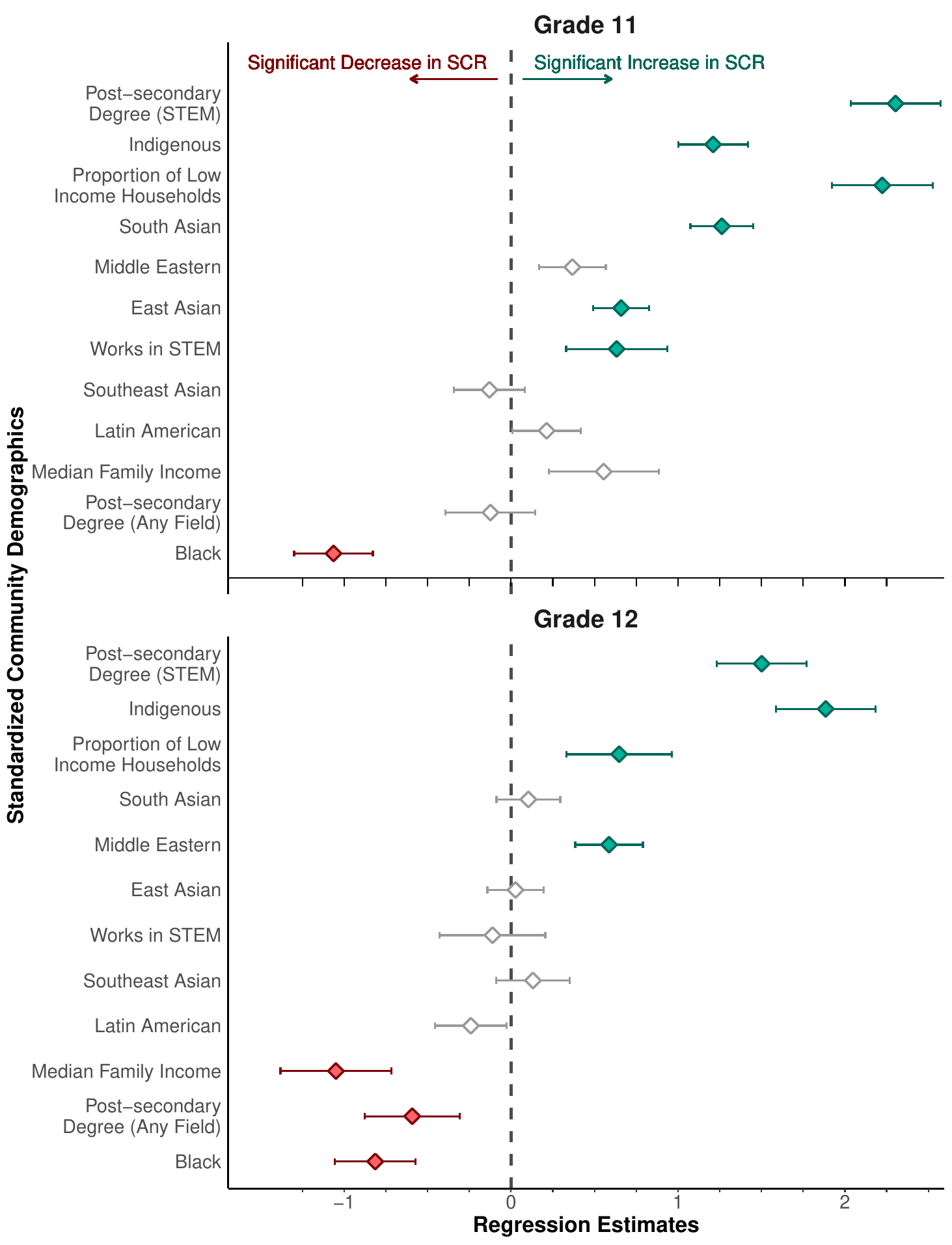}
\caption{Regression estimates for Student Continuation Rates (SCR) into grade 11 (top) and grade 12 physics (bottom), plotted against the different community demographics. Coloured points indicate that regression estimate was found to be statistically significant at a level of $p<.05$; green is for a positive effect while red is for negative effects. Greyed out points were not found to be statistically significant ($p>.05)$. The error bars represent standard error.}
\label{fig:Census_RQ1_Plot}
\end{figure}

Figure \ref{fig:Census_RQ1_Plot} plots the estimates and standard errors calculated by the mixed-effects linear regression model used to address Research Question 1 (Eq. \ref{eq:census_model}). Predictor variables are ordered by the average effect size observed across grades 11 and 12.  Comprehensive regression results for these models can be found in Table \ref{tab:Census-RQ1_Reg} in the appendix. 

Among the predictor variables considered, the largest average effect on SCR is found for the proportion of individuals with a post-secondary degree in STEM ($\beta_{11} = 2.30 \pm 0.27$; $\beta_{12} = 1.50 \pm 0.27$). While this model also controls for the proportion of individuals working in STEM and those with any post-secondary degree, these results suggest important community-level differences between having studied STEM at university and employment in STEM on SCR. Even if one might assume these factors are interconnected, their regression estimates differ considerably. An increased proportion of the community having a post-secondary degree in any field was a statistically significant predictor only for grade 12 physics, with a negative effect ($\beta_{12} = -0.59 \pm 0.29$). Meanwhile, the effect from an increased proportion of the community working in STEM was only statistically significant for grade 11,  predicting an increase in SCR ($\beta_{11} = 0.633 \pm 0.29$).

The proportion of Indigenous individuals within a community had the second-largest average positive effect size ($\beta_{11} = 1.21 \pm 0.21$; $\beta_{12} = 1.89 \pm 0.30$). This was the most influential predictor variable among race or Indigeneity-related community demographics. In comparison, while some other demographic factors connected to race showed positive, statistically significant results, they were less consistent.  Specifically, the predicted effect sizes for the proportions of South Asian ($\beta_{11} = 1.26 \pm 0.19$), Middle Eastern ($\beta_{11} = 0.37 \pm 0.20$), and East Asian individuals ($\beta_{11} = 0.66 \pm 0.17$) were not only smaller but also significant for only one grade level. Southeast Asian and Latin American proportions were not significant in either model. Finally, the proportion of Black individuals within a community was the the sole predictor variable which was negative and statistically significant for both grade 11 ($\beta_{11} = -1.06 \pm 0.24$) and grade 12 physics ($\beta_{12} = -0.81 \pm 0.24$). This stands in stark contrast to all other demographics factors considered.

The final demographic variables considered were those related to family income. The proportion of low-income households was found to have the third largest average effect size, predicting an increase in SCR for both grade 11 and 12 physics ($\beta_{11} = 2.22 \pm 0.30$; $\beta_{12} = 0.65 \pm 0.32$). Median family income of a neighbourhood showed a modest positive estimate for grade 11 ($\beta_{11} = 0.56 \pm 0.33$), but predicted a decline in SCR for grade 12 physics ($\beta_{12} = -1.05 \pm 0.33$).

\subsection{RQ2 - The Gender Gap in SCR}

\begin{figure}
\includegraphics[width = 0.85\linewidth]{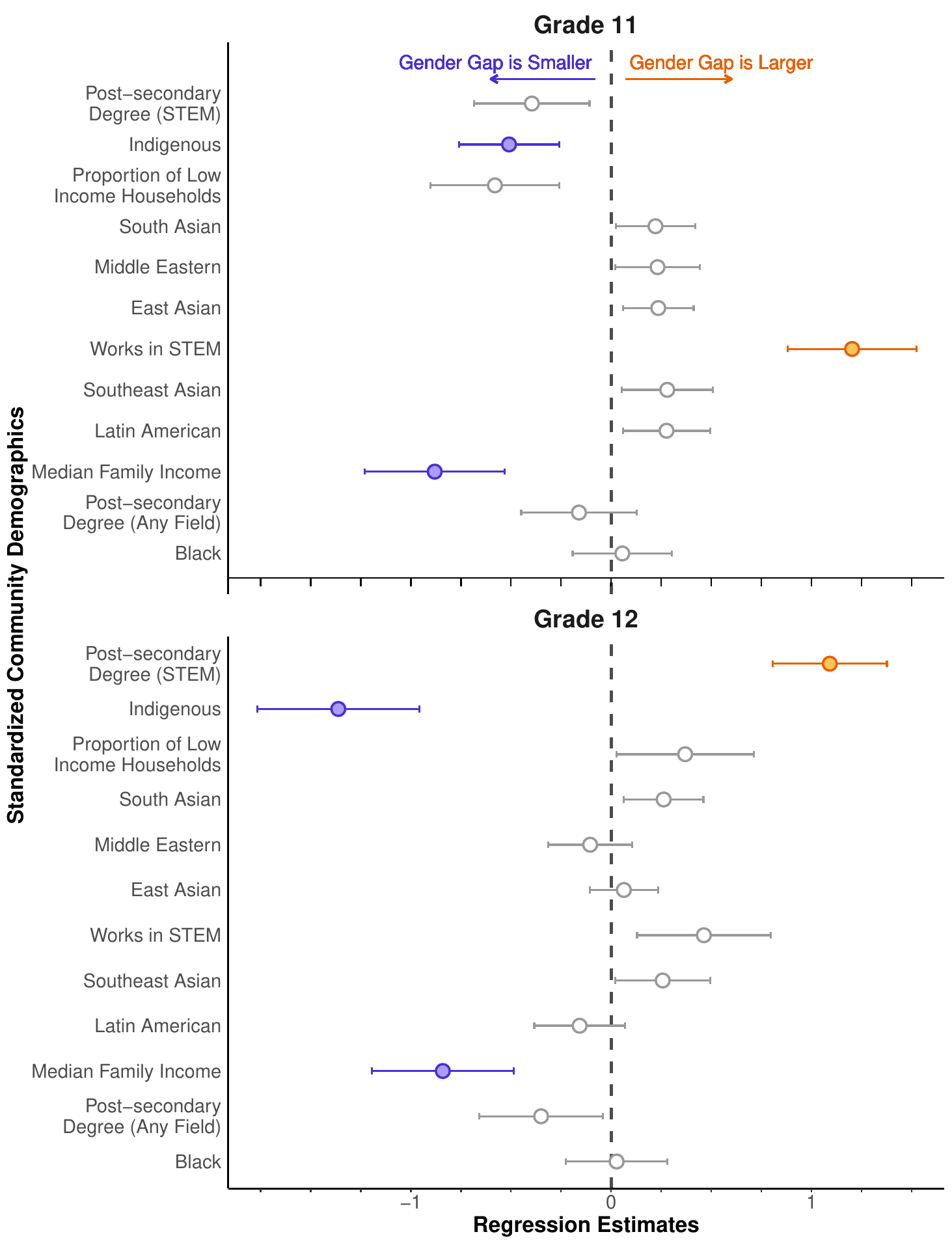}
\caption{Regression estimates for the average gap in Student Continuation Rates (SCR) between male and female students transitioning into grade 11 (top) and grade 12 physics (bottom), plotted against the different community demographics. Coloured points indicate statistically significant regression estimates  $(p<.05)$; orange is for a positive regression estimate, indicating the gender gap between males and females is larger, while purple is for negative regression estimates, indicating that the gender gap is smaller. Greyed out points were not statistically significant ($p>.05)$. The error bars represent the standard error.}
\label{fig:Census_RQ2_Plot}
\end{figure}

The results obtained from the mixed-effects linear regression models for Research Question 2 (\eqref{eq:census_model}) are presented in Figure \ref{fig:Census_RQ2_Plot} while the full regression results are given in Table \ref{tab:Census-RQ2_Reg} within the appendix. To ease comparison, the arrangement of predictor variables in these plots mirror that of Figure \ref{fig:Census_RQ2_Plot}, i.e., sorted based on the average largest effect size found in the  RQ1 models.

In contrast to the RQ1 models, the models for RQ2 have fewer statistically significant regression estimates. In addition, the conditional $R^2$ values for these models are considerably smaller (Tables \ref{tab:Census-RQ1_Reg} and \ref{tab:Census-RQ2_Reg}), suggesting that community-level demographics are much less effective at predicting the gender gap in SCR between male and female students compared to predicting overall variations in SCR for the entire student population.

Of the top three variables for the RQ1 models in terms of average effect sizes — namely, the proportions of individuals with a STEM degree, individuals identifying as Indigenous, and children in low-income households — only the first two were found to have statistically significant estimates for the RQ2 models.  Specifically, communities with a higher percentage of post-secondary STEM graduates showed an larger gender gap in SCR between male students and female students, though only for grade 12 physics $(\beta_{12} = 1.09 \pm 0.29)$.  Conversely, the proportion of Indigenous people in a school’s community predicted a decrease in the gender gap for both grade 11 and 12 physics ($\beta_{11} = -0.51 \pm 0.25$; $\beta_{12} = -1.4 \pm 0.40$). Across the RQ2 models, all other community demographics related to race lacked sufficient evidence to reject the null-hypothesis, i.e. their community-level demographics do not significantly predict the gender gap in SCR. 

For most community demographics related to SES showed little predictive power for RQ2, with the exception of the proportion of individuals working in STEM and median family income. The former predicted an increase in male SCR compared with female SCR, though only for grade 11 physics ($\beta_{11} = 1.20 \pm 0.32$). The latter was found to have negative, statistically significant regression estimates for both grade 11 and 12 physics ($\beta_{11} = -0.88 \pm 0.35$; $\beta_{12} = -0.84 \pm 0.35$). In other words, as median family income increases, while holding all other variables in the model constant, SCR for female students tends to increase relative to that of male students, shrinking the gender gap.

\section{Discussion and Conclusions}\label{sec:census_discuss}

This study used eleven years of comprehensive administrative data from the Ontario Ministry of Education, combined with the Canadian Census, to explore how the demographics of school's surrounding community predict the the rate at which students transition from mandatory grade 10 science to elective physics courses in grades 11 and 12 and the gender gap in continuation rates. Two pairs of mixed-effects linear regression models were built to address these questions.

Our results showed that while many race-based variables in our models were associated with an increase in SCR for grade 11 or 12 physics, the most substantial effect was tied to the percentage of Indigenous people in a school's community. This was a significant predictor of higher SCR in both grades. While other studies have found reduced participation in STEM among Indigenous students \cite{cooper_demographic_2020, polyzou_indigenous_2022, jin_supporting_2021}, our findings diverge from this, potentially due to methodological differences. By incorporating a wider range of socioeconomic factors, we control for many disparities between the Indigenous and non-Indigenous populations in Canada; for example, Indigenous children are twice as likely to live in low-income households \cite{statistics_canada_profile_2023}. This may help to separate the differing effects of Indigeneity from systemic disparities.

In addition, the rise in SCR related to an increase in the Indigenous population within a school's community is larger for female rather than male students. For both grade 11 and 12, schools in communities with more Indigenous residents predicted a smaller gender gap in SCR. During the same time period, other research in Canada has found the growth in post-secondary degree attainment for Indigenous women has significantly outpaced that of Indigenous men \cite{arriagada_achievements_2021}. Future interventions to promote women in physics should capitalize on the momentum of increased SCR for larger Indigenous populations, while also being mindful of the gendered differences in continuation.

In areas with a higher concentration of Black residents, local schools displayed a concerning decrease in SCR for both grade 11 and 12 physics -- the \textit{only} demographic factor to predict a decrease for both grades. Previous research has found that Black students in Ontario’s high schools are disproportionately streamed into non-academic and non-university stream courses, highlighting a systematic bias within Ontario's school system \cite{james_towards_2017}. The observed decrease in SCR is likely another manifestation of these prejudices. 

Furthermore, our RQ2 analysis showed no significant effect on the gender gap in SCR related to the percentage of Black residents in school communities. This finding may seem to contrast with studies highlighting the unique difficulties faced by Black women in physics education \cite{dickens_being_2020, hyater-adams_critical_2018, hyater-adams_deconstructing_2019, mcgee_devalued_2016, rosa_educational_2016}. However, we also find limited evidence for any gender-based gaps in SCR related to the different racial demographics in our model. Thus, though the experience of many racialized women is often markedly different at an individual level, the differences may not be measurable at the school-wide level here, particularly while also controlling for other socioeconomic factors.

Our results also show that parental education significantly influences SCR in physics. As the number of individuals with a post-secondary STEM degree rose, this predicted a large increase in SCR for both grade 11 and 12 physics and the largest average effect observed for RQ1. Conversely, neighbourhoods with more post-secondary degrees in any field predicted a drop in SCR for grade 12 physics. Students seem to follow academic paths mirroring their parents’, suggesting a potential role model effect. When more people are educated in STEM, students in that neighbourhood are more likely to continue into physics. When more people in a neighbourhood are highly educated, but primarily outside STEM (as the number of STEM degrees is held constant), student’s likelihood of continuing into physics goes down. This notion aligns with prior research as children of degree-holding parents are more likely to go to university \cite{turcotte_intergenerational_2011}, but their educational choices are often guided by parental expectations \cite{jacobs_enduring_2006, jones_examining_2022}. 

Similar results were also found in the US where neighbourhoods with more women working in STEM had higher female participation in high school physics \cite{riegle-crumb_gender_2014}.  We also found STEM employment predicted higher SCR, but the results were less definitive. A modest, positive effect was sound for SCR into grade 11, it was absent for grade 12. In addition, our findings found that neighbourhoods with more people educated or working in STEM predicted a wider male-female continuation gap. While controlling for the level of parental education, the work by \cite{riegle-crumb_gender_2014} did not account for the specific field of study. A report from the American Institute of Physics found only 4\% of new physics bachelor's degree holders working in the private sector worked directly in physics and astronomy \cite{aip_statistics_initial_2022}. Many transition to other STEM fields or leave STEM entirely. Despite most physics graduates not directly working in that field, we posit that many parents will still promote it as a viable path for their children to achieve career success.

Finally, in neighbourhoods with more children in low-income households, we observed higher physics continuation rates, even when adjusting for median family income. Moreover, in such neighbourhoods, the gender gap in continuation between male and female students decreased for grade 11 physics. This aligns with US based research showing lower SES students often select their field of study based on primarily economic concerns \cite{brand_who_2010} and are more confident in choosing STEM majors than higher SES students \cite{george-jackson_college_2012}. Globally, less gender-equal countries often see more women in STEM, possibly viewing it as a more secure path to financial independence   \cite{stoet_gender-equality_2018}. Similar mechanisms may be driving the effects seen in our analysis.

Future work can build upon these results in two main ways. First, this study primarily offers descriptive insights, highlighting the correlations between community-level demographics and student continuation into high school physics without establishing causality or delving into the reasons behind these patterns. Studies focusing on individual students, especially those exploring the role of demographics on affective metrics like Physics Identity  \cite{hazari_connecting_2010}, can provide a deeper understanding, going beyond just observing disparities in engagement as done in this research. Additionally, even with the community-level analysis conducted here, our models  have several significant findings related to the intersectionality of gender and other demographics. That these intersectional effects are discernible at a school-wide level, underscores the need for researchers and policymakers to be acutely aware of such differences. Understanding these nuances is essential if we hope to fully address the gender gap in physics.

\section*{Author Statements}
\subsection*{acknowledgments}
We wish to acknowledge the support of Leo Custode for helpful discussions about the statistical analysis in this work. This work was funded in part by the Social Science and Humanities Research Council of Canada through a Partnership Grant (895-2017-1025).
\subsection*{Competing Interests}
The authors declare there are no competing interests. 
\subsection*{Author Contributions}
EC obtained, analyzed, and interpreted the enrolment data from the OME as well as contributed to the writing and editing of the manuscript. MAW and MW provided guidance to the direction of analysis and the research questions we sought to answer as well as contributed to the writing and editing of the manuscript. All authors read and approved the final manuscript.
\subsection*{Funding}
Funding for this work was provided by a Partnership Grant from the Social Sciences and Humanities Research Council of Canada (\#895-2017-1025) as well as a Doctoral Fellowship from the Social Sciences and Humanities Research Council of Canada (\#752-2020-1617).
\subsection*{Availability of Data}
The data that support the findings of this study are available from the Ontario Ministry of Education, but restrictions apply to the availability of these data, which were used under license for the current study, and so are not publicly available. Data are however available from the authors upon reasonable request and with permission of the OME.

\appendix

\section{Full Regression Tables}

% Table created by stargazer v.5.2.3 by Marek Hlavac, Social Policy Institute. E-mail: marek.hlavac at gmail.com
% Date and time: Mon, Oct 09, 2023 - 08:56:38
\begin{table}[!htbp] \centering 
  \caption[Full regression results for Census paper RQ1]{Detailed regression results for the mixed-effects regression model comparing total SCR into grade 11 and 12 physics versus community-level demographics.} 
  \label{tab:Census-RQ1_Reg} 
\begin{tabular*}{\textwidth}{@{\extracolsep{\fill}} l D{.}{.}{4} D{.}{.}{3} D{.}{.}{3} c D{.}{.}{4} D{.}{.}{3} D{.}{.}{3}} 
\toprule
 & \multicolumn{3}{c}{Physics 11} && \multicolumn{3}{c}{Physics 12}\\
\cline{2-4} \cline{6-8}
Community Demographics & \multicolumn{1}{c}{Estimate} & \multicolumn{1}{c}{(SE)} & \multicolumn{1}{c}{p-value} && \multicolumn{1}{c}{Estimate} & \multicolumn{1}{c}{(SE)} & \multicolumn{1}{c}{p-value} \\
\midrule
 Black & -1.063^{***} & (0.236) & <.001 && -0.814^{***} & (0.242) & <.001 \\ 
  \\ 
 
East Asian & 0.659^{***} & (0.169) & <.001 && 0.027 & (0.168) & .874 \\ 
  \\ 
 
Southeast Asian & -0.129 & (0.212) & .543 && 0.131 & (0.220) & .550 \\ 
  \\ 
 
South Asian & 1.263^{***} & (0.189) & <.001 && 0.104 & (0.191) & .586 \\ 
  \\ 
 
Middle Eastern & 0.368^{.} & (0.200) & .066 && 0.587^{**} & (0.203) & .004 \\ 
  \\ 
 
Latin American & 0.213 & (0.205) & .299 && -0.241 & (0.214) & .260 \\ 
  \\ 
 
Indigenous & 1.210^{***} & (0.209) & <.001 && 1.885^{***} & (0.298) & <.001 \\ 
  \\ 
 
Post-Secondary Degrees (Any Field) & -0.124 & (0.269) & .646 && -0.592^{*} & (0.285) & .038 \\ 
\\
 
Post-Secondary Degrees (STEM) & 2.304^{***} & (0.268) & <.001 && 1.501^{***} & (0.270) & <.001 \\ 
\\ 
 
People Working in STEM & 0.632^{*} & (0.303) & .037 && -0.111 & (0.317) & .725 \\ 
\\ 
 
\multirow{2}{6cm}{Proportion of Children in Low-Income Households} & 2.224^{***} & (0.303) & <.001 && 0.647^{*} & (0.317) & .042 \\ 
\\ 
 
Median Family Income & 0.555^{.} & (0.329) & .093 && -1.049^{**} & (0.333) & .002 \\  
\midrule
Observations & \multicolumn{3}{c}{11,179} && \multicolumn{3}{c}{7,579} \\ 
Conditional $R^2$ & \multicolumn{3}{c}{0.122} && \multicolumn{3}{c}{0.105} \\
\bottomrule
\multicolumn{8}{r}{\textit{Note:} $^{*}$p$<$.05; $^{**}$p$<$.01; $^{***}$p$<$.001} \\ 
\end{tabular*} 
\end{table}

\begin{table}[!htbp] \centering 
  \caption[Full regression results for Census paper RQ1]{Detailed regression results for the mixed-effects regression model comparing the male-female gender gap in SCR for grade 11 and 12 physics versus community-level demographics.} 
  \label{tab:Census-RQ2_Reg} 
\begin{tabular*}{\textwidth}{@{\extracolsep{\fill}} l D{.}{.}{4} D{.}{.}{3} D{.}{.}{3} c D{.}{.}{4} D{.}{.}{3} D{.}{.}{3}} 
\toprule
 & \multicolumn{3}{c}{Physics 11} && \multicolumn{3}{c}{Physics 12}\\
\cline{2-4} \cline{6-8}
Community Demographics & \multicolumn{1}{c}{Estimate} & \multicolumn{1}{c}{(SE)} & \multicolumn{1}{c}{p-value} && \multicolumn{1}{c}{Estimate} & \multicolumn{1}{c}{(SE)} & \multicolumn{1}{c}{p-value} \\
\midrule
Black & 0.056 & (0.248) & .823 && 0.027 & (0.253) & .915 \\ 
\\
East Asian & 0.235 & (0.176) & .182 && 0.063 & (0.171) & .711 \\ 
\\
Southeast Asian & 0.280 & (0.227) && .219 & 0.257 & (0.236) & .277 \\ 
\\
South Asian & 0.221 & (0.198) & .264 && 0.262 & (0.199) & .188 \\ 
\\
Middle Eastern & 0.231 & (0.211) & .272 && -0.105 & (0.210) & .616 \\ 
\\
Latin American & 0.276 & (0.218) & .205 && -0.158 & (0.227) & .486 \\ 
\\
Indigenous & -0.510^{*} & (0.250) & .042 && -1.362^{***} & (0.405) & <.001 \\ 
\\
Post-Secondary Degrees (Any Field) & -0.161 & (0.288) & .577 && -0.350 & (0.308) & .256 \\ 
\\
Post-Secondary Degrees (STEM) & -0.396 & (0.288) & .170 && 1.091^{***} & (0.285) & <.001 \\ 
\\
People Working in STEM & 1.203^{***} & (0.322) & <.001 && 0.463 & (0.333) & .166 \\ 
\\
\multirow{2}{6cm}{Proportion of Children in Low-Income Households} & -0.581 & (0.322) & .072 && 0.369 & (0.344) & .283 \\ 
\\
Median Family Income & -0.881^{*} & (0.349) & .012 && -0.841^{*} & (0.354) & .018 \\  
\midrule
Observations & \multicolumn{3}{c}{11,179} && \multicolumn{3}{c}{7,579} \\ 
Conditional $R^2$ & \multicolumn{3}{c}{0.122} && \multicolumn{3}{c}{0.105} \\
\bottomrule
\multicolumn{8}{r}{\textit{Note:} $^{*}$p$<$.05; $^{**}$p$<$.01; $^{***}$p$<$.001} \\ 
\end{tabular*} 
\end{table}

\newpage

\bibliography{references} %Produces the bibliography via BibTeX.

\begin{thebibliography}{55}%
\makeatletter
\providecommand \@ifxundefined [1]{%
 \@ifx{#1\undefined}
}%
\providecommand \@ifnum [1]{%
 \ifnum #1\expandafter \@firstoftwo
 \else \expandafter \@secondoftwo
 \fi
}%
\providecommand \@ifx [1]{%
 \ifx #1\expandafter \@firstoftwo
 \else \expandafter \@secondoftwo
 \fi
}%
\providecommand \natexlab [1]{#1}%
\providecommand \enquote  [1]{``#1''}%
\providecommand \bibnamefont  [1]{#1}%
\providecommand \bibfnamefont [1]{#1}%
\providecommand \citenamefont [1]{#1}%
\providecommand \href@noop [0]{\@secondoftwo}%
\providecommand \href [0]{\begingroup \@sanitize@url \@href}%
\providecommand \@href[1]{\@@startlink{#1}\@@href}%
\providecommand \@@href[1]{\endgroup#1\@@endlink}%
\providecommand \@sanitize@url [0]{\catcode `\\12\catcode `\$12\catcode `\&12\catcode `\#12\catcode `\^12\catcode `\_12\catcode `\%12\relax}%
\providecommand \@@startlink[1]{}%
\providecommand \@@endlink[0]{}%
\providecommand \url  [0]{\begingroup\@sanitize@url \@url }%
\providecommand \@url [1]{\endgroup\@href {#1}{\urlprefix }}%
\providecommand \urlprefix  [0]{URL }%
\providecommand \Eprint [0]{\href }%
\providecommand \doibase [0]{https://doi.org/}%
\providecommand \selectlanguage [0]{\@gobble}%
\providecommand \bibinfo  [0]{\@secondoftwo}%
\providecommand \bibfield  [0]{\@secondoftwo}%
\providecommand \translation [1]{[#1]}%
\providecommand \BibitemOpen [0]{}%
\providecommand \bibitemStop [0]{}%
\providecommand \bibitemNoStop [0]{.\EOS\space}%
\providecommand \EOS [0]{\spacefactor3000\relax}%
\providecommand \BibitemShut  [1]{\csname bibitem#1\endcsname}%
\let\auto@bib@innerbib\@empty
%</preamble>
\bibitem [{\citenamefont {APS}\ and\ \citenamefont {IPEDS}(2022)}]{aps_bachelors_2022}%
  \BibitemOpen
  \bibfield  {author} {\bibinfo {author} {\bibnamefont {APS}}\ and\ \bibinfo {author} {\bibnamefont {IPEDS}},\ }\href {http://www.aps.org/programs/education/statistics/womenmajors.cfm} {\bibinfo {title} {Bachelor’s degrees earned by women, by major}} (\bibinfo {year} {2022})\BibitemShut {NoStop}%
\bibitem [{\citenamefont {Huang}\ \emph {et~al.}(2020)\citenamefont {Huang}, \citenamefont {Gates}, \citenamefont {Sinatra},\ and\ \citenamefont {Barabási}}]{huang_historical_2020}%
  \BibitemOpen
  \bibfield  {author} {\bibinfo {author} {\bibfnamefont {J.}~\bibnamefont {Huang}}, \bibinfo {author} {\bibfnamefont {A.~J.}\ \bibnamefont {Gates}}, \bibinfo {author} {\bibfnamefont {R.}~\bibnamefont {Sinatra}},\ and\ \bibinfo {author} {\bibfnamefont {A.-L.}\ \bibnamefont {Barabási}},\ }\bibfield  {title} {\bibinfo {title} {Historical comparison of gender inequality in scientific careers across countries and disciplines},\ }\href {https://doi.org/10.1073/pnas.1914221117} {\bibfield  {journal} {\bibinfo  {journal} {Proceedings of the National Academy of Sciences}\ }\textbf {\bibinfo {volume} {117}},\ \bibinfo {pages} {4609} (\bibinfo {year} {2020})}\BibitemShut {NoStop}%
\bibitem [{\citenamefont {Krakehl}\ and\ \citenamefont {Kelly}(2021)}]{krakehl_intersectional_2021}%
  \BibitemOpen
  \bibfield  {author} {\bibinfo {author} {\bibfnamefont {R.}~\bibnamefont {Krakehl}}\ and\ \bibinfo {author} {\bibfnamefont {A.~M.}\ \bibnamefont {Kelly}},\ }\bibfield  {title} {\bibinfo {title} {Intersectional analysis of {Advanced} {Placement} {Physics} participation and performance by gender and ethnicity},\ }\href {https://doi.org/10.1103/PhysRevPhysEducRes.17.020105} {\bibfield  {journal} {\bibinfo  {journal} {Physical Review Physics Education Research}\ }\textbf {\bibinfo {volume} {17}},\ \bibinfo {pages} {020105} (\bibinfo {year} {2021})},\ \bibinfo {note} {publisher: American Physical Society}\BibitemShut {NoStop}%
\bibitem [{\citenamefont {Hazari}\ \emph {et~al.}(2010)\citenamefont {Hazari}, \citenamefont {Sonnert}, \citenamefont {Sadler},\ and\ \citenamefont {Shanahan}}]{hazari_connecting_2010}%
  \BibitemOpen
  \bibfield  {author} {\bibinfo {author} {\bibfnamefont {Z.}~\bibnamefont {Hazari}}, \bibinfo {author} {\bibfnamefont {G.}~\bibnamefont {Sonnert}}, \bibinfo {author} {\bibfnamefont {P.~M.}\ \bibnamefont {Sadler}},\ and\ \bibinfo {author} {\bibfnamefont {M.-C.}\ \bibnamefont {Shanahan}},\ }\bibfield  {title} {{\selectlanguage {en}\bibinfo {title} {Connecting high school physics experiences, outcome expectations, physics identity, and physics career choice: {A} gender study}},\ }\href {https://doi.org/10.1002/tea.20363} {\bibfield  {journal} {\bibinfo  {journal} {Journal of Research in Science Teaching}\ }\textbf {\bibinfo {volume} {47}},\ \bibinfo {pages} {978} (\bibinfo {year} {2010})}\BibitemShut {NoStop}%
\bibitem [{\citenamefont {Card}\ and\ \citenamefont {Payne}(2021)}]{card_high_2021}%
  \BibitemOpen
  \bibfield  {author} {\bibinfo {author} {\bibfnamefont {D.}~\bibnamefont {Card}}\ and\ \bibinfo {author} {\bibfnamefont {A.~A.}\ \bibnamefont {Payne}},\ }\bibfield  {title} {{\selectlanguage {en}\bibinfo {title} {High {School} {Choices} and the {Gender} {Gap} in {Stem}}},\ }\href {https://doi.org/10.1111/ecin.12934} {\bibfield  {journal} {\bibinfo  {journal} {Economic Inquiry}\ }\textbf {\bibinfo {volume} {59}},\ \bibinfo {pages} {9} (\bibinfo {year} {2021})}\BibitemShut {NoStop}%
\bibitem [{\citenamefont {Corrigan}\ \emph {et~al.}(2023)\citenamefont {Corrigan}, \citenamefont {Williams},\ and\ \citenamefont {Wells}}]{corrigan_high_2023}%
  \BibitemOpen
  \bibfield  {author} {\bibinfo {author} {\bibfnamefont {E.}~\bibnamefont {Corrigan}}, \bibinfo {author} {\bibfnamefont {M.}~\bibnamefont {Williams}},\ and\ \bibinfo {author} {\bibfnamefont {M.~A.}\ \bibnamefont {Wells}},\ }\bibfield  {title} {{\selectlanguage {en}\bibinfo {title} {High {School} {Enrolment} {Choices}—{Understanding} the {STEM} {Gender} {Gap}}},\ }\bibfield  {journal} {\bibinfo  {journal} {Canadian Journal of Science, Mathematics and Technology Education}\ }\href {https://doi.org/10.1007/s42330-023-00285-y} {10.1007/s42330-023-00285-y} (\bibinfo {year} {2023})\BibitemShut {NoStop}%
\bibitem [{\citenamefont {Hennessey}\ \emph {et~al.}(2019)\citenamefont {Hennessey}, \citenamefont {Cole}, \citenamefont {Shastri}, \citenamefont {Esquivel}, \citenamefont {Singh}, \citenamefont {Johnson},\ and\ \citenamefont {Ghose}}]{hennessey_workshop_2019}%
  \BibitemOpen
  \bibfield  {author} {\bibinfo {author} {\bibfnamefont {E.}~\bibnamefont {Hennessey}}, \bibinfo {author} {\bibfnamefont {J.}~\bibnamefont {Cole}}, \bibinfo {author} {\bibfnamefont {P.}~\bibnamefont {Shastri}}, \bibinfo {author} {\bibfnamefont {J.}~\bibnamefont {Esquivel}}, \bibinfo {author} {\bibfnamefont {C.}~\bibnamefont {Singh}}, \bibinfo {author} {\bibfnamefont {R.}~\bibnamefont {Johnson}},\ and\ \bibinfo {author} {\bibfnamefont {S.}~\bibnamefont {Ghose}},\ }\bibfield  {title} {\bibinfo {title} {Workshop report: {Intersecting} identities—gender and intersectionality in physics},\ }\href {https://doi.org/10.1063/1.5110070} {\bibfield  {journal} {\bibinfo  {journal} {AIP Conference Proceedings}\ }\textbf {\bibinfo {volume} {2109}},\ \bibinfo {pages} {040001} (\bibinfo {year} {2019})},\ \bibinfo {note} {publisher: American Institute of Physics}\BibitemShut {NoStop}%
\bibitem [{\citenamefont {Bécares}\ and\ \citenamefont {Priest}(2015)}]{becares_understanding_2015}%
  \BibitemOpen
  \bibfield  {author} {\bibinfo {author} {\bibfnamefont {L.}~\bibnamefont {Bécares}}\ and\ \bibinfo {author} {\bibfnamefont {N.}~\bibnamefont {Priest}},\ }\bibfield  {title} {{\selectlanguage {en}\bibinfo {title} {Understanding the influence of race/ethnicity, gender, and class on inequalities in academic and non-academic outcomes among eighth-grade students: findings from an intersectionality approach}},\ }\href {https://doi.org/10.1371/journal.pone.0141363} {\bibfield  {journal} {\bibinfo  {journal} {PLOS ONE}\ }\textbf {\bibinfo {volume} {10}},\ \bibinfo {pages} {e0141363} (\bibinfo {year} {2015})},\ \bibinfo {note} {publisher: Public Library of Science}\BibitemShut {NoStop}%
\bibitem [{\citenamefont {Society}(2020)}]{american_physical_society_physics_2020}%
  \BibitemOpen
  \bibfield  {author} {\bibinfo {author} {\bibfnamefont {A.~P.}\ \bibnamefont {Society}},\ }\href {https://www.aps.org/programs/education/statistics/degreesbyrace.cfm} {{\selectlanguage {en}\bibinfo {title} {Physics degrees by race/ethnicity}}} (\bibinfo {year} {2020})\BibitemShut {NoStop}%
\bibitem [{\citenamefont {Smolina}\ \emph {et~al.}(2021)\citenamefont {Smolina}, \citenamefont {Hewitt}, \citenamefont {Ghose}, \citenamefont {Hennessey}, \citenamefont {Tassone},\ and\ \citenamefont {Hennessey}}]{smolina_can_2021}%
  \BibitemOpen
  \bibfield  {author} {\bibinfo {author} {\bibfnamefont {A.}~\bibnamefont {Smolina}}, \bibinfo {author} {\bibfnamefont {K.}~\bibnamefont {Hewitt}}, \bibinfo {author} {\bibfnamefont {S.}~\bibnamefont {Ghose}}, \bibinfo {author} {\bibfnamefont {E.}~\bibnamefont {Hennessey}}, \bibinfo {author} {\bibfnamefont {A.}~\bibnamefont {Tassone}},\ and\ \bibinfo {author} {\bibfnamefont {S.}~\bibnamefont {Hennessey}},\ }\href {https://www.canphyscounts.ca/survey-data} {{\selectlanguage {en}\emph {\bibinfo {title} {Can {Phys} {Counts}: the first {Canada}-wide equity, diversity, and inclusions in physics survey}}}},\ \bibinfo {type} {Tech. Rep.}\ (\bibinfo  {institution} {Canadian Association of Physicists},\ \bibinfo {year} {2021})\BibitemShut {NoStop}%
\bibitem [{\citenamefont {Espinosa}(2009)}]{espinosa_pipelines_2009}%
  \BibitemOpen
  \bibfield  {author} {\bibinfo {author} {\bibfnamefont {L.~L.}\ \bibnamefont {Espinosa}},\ }{\selectlanguage {English}\emph {\bibinfo {title} {Pipelines and pathways: {Women} of color in {STEM} majors and the experiences that shape their persistence}}},\ \href {https://www.proquest.com/docview/304854082/abstract/84A8F91A703E4786PQ/1} {\bibinfo {type} {Ph.{D}.}},\ \bibinfo  {school} {University of California, Los Angeles}, \bibinfo {address} {United States -- California} (\bibinfo {year} {2009}),\ \bibinfo {note} {iSBN: 9781109622300}\BibitemShut {NoStop}%
\bibitem [{\citenamefont {Ma}\ and\ \citenamefont {Liu}(2015)}]{ma_race_2015}%
  \BibitemOpen
  \bibfield  {author} {\bibinfo {author} {\bibfnamefont {Y.}~\bibnamefont {Ma}}\ and\ \bibinfo {author} {\bibfnamefont {Y.}~\bibnamefont {Liu}},\ }\bibfield  {title} {{\selectlanguage {en}\bibinfo {title} {Race and {STEM} {Degree} {Attainment}}},\ }\href {https://doi.org/10.1111/soc4.12274} {\bibfield  {journal} {\bibinfo  {journal} {Sociology Compass}\ }\textbf {\bibinfo {volume} {9}},\ \bibinfo {pages} {609} (\bibinfo {year} {2015})}\BibitemShut {NoStop}%
\bibitem [{\citenamefont {Polyzou}(2022)}]{polyzou_indigenous_2022}%
  \BibitemOpen
  \bibfield  {author} {\bibinfo {author} {\bibfnamefont {C.}~\bibnamefont {Polyzou}},\ }\href {https://engineerscanada.ca/reports/indigenous-inclusion-in-engineering} {{\selectlanguage {en}\emph {\bibinfo {title} {Indigenous {Inclusion} in {Engineering}}}}},\ \bibinfo {type} {Tech. Rep.}\ (\bibinfo  {institution} {Engineers Canada},\ \bibinfo {year} {2022})\BibitemShut {NoStop}%
\bibitem [{\citenamefont {Eaton}\ \emph {et~al.}(2019)\citenamefont {Eaton}, \citenamefont {Saunders}, \citenamefont {Jacobson},\ and\ \citenamefont {West}}]{eaton_how_2019}%
  \BibitemOpen
  \bibfield  {author} {\bibinfo {author} {\bibfnamefont {A.~A.}\ \bibnamefont {Eaton}}, \bibinfo {author} {\bibfnamefont {J.~F.}\ \bibnamefont {Saunders}}, \bibinfo {author} {\bibfnamefont {R.~K.}\ \bibnamefont {Jacobson}},\ and\ \bibinfo {author} {\bibfnamefont {K.}~\bibnamefont {West}},\ }\bibfield  {title} {\bibinfo {title} {How {Gender} and {Race} {Stereotypes} {Impact} the {Advancement} of {Scholars} in {STEM}: {Professors}’ {Biased} {Evaluations} of {Physics} and {Biology} {Post}-{Doctoral} {Candidates}},\ }\href {https://doi.org/10.1007/s11199-019-01052-w} {\bibfield  {journal} {\bibinfo  {journal} {Sex Roles}\ }\textbf {\bibinfo {volume} {82}},\ \bibinfo {pages} {127} (\bibinfo {year} {2019})},\ \bibinfo {note} {publisher: Springer US}\BibitemShut {NoStop}%
\bibitem [{\citenamefont {Arriagada}(2021)}]{arriagada_achievements_2021}%
  \BibitemOpen
  \bibfield  {author} {\bibinfo {author} {\bibfnamefont {P.}~\bibnamefont {Arriagada}},\ }\href {https://www150.statcan.gc.ca/n1/en/pub/75-006-x/2021001/article/00009-eng.pdf?st=tA78DsCp} {\emph {\bibinfo {title} {The achievements, experiences and labour market outcomes of first nations, métis and inuit women with bachelor’s degrees or higher}}},\ \bibinfo {type} {Tech. Rep.}\ \bibinfo {number} {75-006-X}\ (\bibinfo  {institution} {Statistics Canada},\ \bibinfo {year} {2021})\BibitemShut {NoStop}%
\bibitem [{\citenamefont {Walpole}(2003)}]{walpole_socioeconomic_2003}%
  \BibitemOpen
  \bibfield  {author} {\bibinfo {author} {\bibfnamefont {M.}~\bibnamefont {Walpole}},\ }\bibfield  {title} {\bibinfo {title} {Socioeconomic {Status} and {College}: {How} {SES} {Affects} {College} {Experiences} and {Outcomes}},\ }\href {https://doi.org/10.1353/rhe.2003.0044} {\bibfield  {journal} {\bibinfo  {journal} {The Review of Higher Education}\ }\textbf {\bibinfo {volume} {27}},\ \bibinfo {pages} {45} (\bibinfo {year} {2003})},\ \bibinfo {note} {publisher: Johns Hopkins University Press}\BibitemShut {NoStop}%
\bibitem [{\citenamefont {Frenette}(2017)}]{frenette_postsecondary_2017}%
  \BibitemOpen
  \bibfield  {author} {\bibinfo {author} {\bibfnamefont {M.}~\bibnamefont {Frenette}},\ }\href@noop {} {{\selectlanguage {en}\emph {\bibinfo {title} {Postsecondary enrolment by parental income: recent national and provincial trends}}}}\ (\bibinfo  {publisher} {Statistics Canada},\ \bibinfo {year} {2017})\ \bibinfo {note} {oCLC: 1011505358}\BibitemShut {NoStop}%
\bibitem [{\citenamefont {Brand}\ and\ \citenamefont {Xie}(2010)}]{brand_who_2010}%
  \BibitemOpen
  \bibfield  {author} {\bibinfo {author} {\bibfnamefont {J.~E.}\ \bibnamefont {Brand}}\ and\ \bibinfo {author} {\bibfnamefont {Y.}~\bibnamefont {Xie}},\ }\bibfield  {title} {{\selectlanguage {en}\bibinfo {title} {Who benefits most from college?: evidence for negative selection in heterogeneous economic returns to higher education}},\ }\bibfield  {journal} {\bibinfo  {journal} {American Sociological Review}\ }\href {https://doi.org/10.1177/0003122410363567} {10.1177/0003122410363567} (\bibinfo {year} {2010}),\ \bibinfo {note} {publisher: SAGE PublicationsSage CA: Los Angeles, CA}\BibitemShut {NoStop}%
\bibitem [{\citenamefont {Lichtenberger}\ and\ \citenamefont {George-Jackson}(2012)}]{lichtenberger_predicting_2012}%
  \BibitemOpen
  \bibfield  {author} {\bibinfo {author} {\bibfnamefont {E.}~\bibnamefont {Lichtenberger}}\ and\ \bibinfo {author} {\bibfnamefont {C.}~\bibnamefont {George-Jackson}},\ }\bibfield  {title} {{\selectlanguage {en}\bibinfo {title} {Predicting {High} {School} {Students}’ {Interest} in {Majoring} in a {STEM} {Field}: {Insight} into {High} {School} {Students}’ {Postsecondary} {Plans}}},\ }\bibfield  {journal} {\bibinfo  {journal} {Journal of Career and Technical Education}\ }\textbf {\bibinfo {volume} {28}},\ \href {https://doi.org/10.21061/jcte.v28i1.571} {10.21061/jcte.v28i1.571} (\bibinfo {year} {2012})\BibitemShut {NoStop}%
\bibitem [{\citenamefont {Cataldi}\ \emph {et~al.}(2018)\citenamefont {Cataldi}, \citenamefont {Bennett}, \citenamefont {Chen},\ and\ \citenamefont {Simone}}]{cataldi_first-generation_2018}%
  \BibitemOpen
  \bibfield  {author} {\bibinfo {author} {\bibfnamefont {E.~F.}\ \bibnamefont {Cataldi}}, \bibinfo {author} {\bibfnamefont {C.~T.}\ \bibnamefont {Bennett}}, \bibinfo {author} {\bibfnamefont {X.}~\bibnamefont {Chen}},\ and\ \bibinfo {author} {\bibfnamefont {S.~A.}\ \bibnamefont {Simone}},\ }\href@noop {} {{\selectlanguage {en}\emph {\bibinfo {title} {First-{Generation} {Students}: {College} {Access}, {Persistence}, and {Postbachelor}’s {Outcomes}}}}},\ \bibinfo {type} {Tech. Rep.}\ (\bibinfo  {institution} {U.S. Department of Education: National Center for Education Statistics},\ \bibinfo {year} {2018})\BibitemShut {NoStop}%
\bibitem [{\citenamefont {Jacobs}\ \emph {et~al.}(2006)\citenamefont {Jacobs}, \citenamefont {Chhin},\ and\ \citenamefont {Bleeker}}]{jacobs_enduring_2006}%
  \BibitemOpen
  \bibfield  {author} {\bibinfo {author} {\bibfnamefont {J.~E.}\ \bibnamefont {Jacobs}}, \bibinfo {author} {\bibfnamefont {C.~S.}\ \bibnamefont {Chhin}},\ and\ \bibinfo {author} {\bibfnamefont {M.~M.}\ \bibnamefont {Bleeker}},\ }\bibfield  {title} {\bibinfo {title} {Enduring links: {Parents}' expectations and their young adult children's gender-typed occupational choices},\ }\href {https://doi.org/10.1080/13803610600765851} {\bibfield  {journal} {\bibinfo  {journal} {Educational Research and Evaluation}\ }\textbf {\bibinfo {volume} {12}},\ \bibinfo {pages} {395} (\bibinfo {year} {2006})},\ \bibinfo {note} {publisher: Routledge \_eprint: https://doi.org/10.1080/13803610600765851}\BibitemShut {NoStop}%
\bibitem [{\citenamefont {Jones}\ and\ \citenamefont {Hamer}(2022)}]{jones_examining_2022}%
  \BibitemOpen
  \bibfield  {author} {\bibinfo {author} {\bibfnamefont {K.~L.}\ \bibnamefont {Jones}}\ and\ \bibinfo {author} {\bibfnamefont {J.~M.~M.}\ \bibnamefont {Hamer}},\ }\bibfield  {title} {\bibinfo {title} {Examining the relationship between parent/carer’s attitudes, beliefs and their child’s future participation in physics},\ }\href {https://doi.org/10.1080/09500693.2021.2021457} {\bibfield  {journal} {\bibinfo  {journal} {International Journal of Science Education}\ }\textbf {\bibinfo {volume} {44}},\ \bibinfo {pages} {201} (\bibinfo {year} {2022})},\ \bibinfo {note} {publisher: Routledge \_eprint: https://doi.org/10.1080/09500693.2021.2021457}\BibitemShut {NoStop}%
\bibitem [{\citenamefont {Riegle-Crumb}\ and\ \citenamefont {Moore}(2014)}]{riegle-crumb_gender_2014}%
  \BibitemOpen
  \bibfield  {author} {\bibinfo {author} {\bibfnamefont {C.}~\bibnamefont {Riegle-Crumb}}\ and\ \bibinfo {author} {\bibfnamefont {C.}~\bibnamefont {Moore}},\ }\bibfield  {title} {{\selectlanguage {en}\bibinfo {title} {The {Gender} {Gap} in {High} {School} {Physics}: {Considering} the {Context} of {Local} {Communities}: {The} {Gender} {Gap} in {High} {School} {Physics}}},\ }\href {https://doi.org/10.1111/ssqu.12022} {\bibfield  {journal} {\bibinfo  {journal} {Social Science Quarterly}\ }\textbf {\bibinfo {volume} {95}},\ \bibinfo {pages} {253} (\bibinfo {year} {2014})}\BibitemShut {NoStop}%
\bibitem [{\citenamefont {{Government of Ontario}}(2022{\natexlab{a}})}]{government_of_ontario_census_2022-2}%
  \BibitemOpen
  \bibfield  {author} {\bibinfo {author} {\bibnamefont {{Government of Ontario}}},\ }\href {http://www.ontario.ca/document/2016-census-highlights/fact-sheet-2-population-counts-census-subdivisions-csds-ontario} {{\selectlanguage {en}\bibinfo {title} {Census {Fact} {Sheet} 2: {Population} counts: {Census} {Subdivisions} ({CSDs}) in {Ontario} {\textbar} 2016 census highlights}}} (\bibinfo {year} {2022}{\natexlab{a}})\BibitemShut {NoStop}%
\bibitem [{\citenamefont {{Government of Ontario}}(2022{\natexlab{b}})}]{government_of_ontario_census_2022}%
  \BibitemOpen
  \bibfield  {author} {\bibinfo {author} {\bibnamefont {{Government of Ontario}}},\ }\href {http://www.ontario.ca/document/2016-census-highlights/fact-sheet-9-ethnic-origin-and-visible-minorities} {{\selectlanguage {en}\bibinfo {title} {Census {Fact} {Sheet} 9: {Ethnic} origin and visible minorities {\textbar} 2016 census highlights}}} (\bibinfo {year} {2022}{\natexlab{b}})\BibitemShut {NoStop}%
\bibitem [{\citenamefont {{Government of Ontario}}(2022{\natexlab{c}})}]{government_of_ontario_census_2022-1}%
  \BibitemOpen
  \bibfield  {author} {\bibinfo {author} {\bibnamefont {{Government of Ontario}}},\ }\href {http://www.ontario.ca/document/2016-census-highlights/fact-sheet-7-income} {{\selectlanguage {en}\bibinfo {title} {Census {Fact} {Sheet} 7: {Income} {\textbar} 2016 census highlights}}} (\bibinfo {year} {2022}{\natexlab{c}})\BibitemShut {NoStop}%
\bibitem [{\citenamefont {{Government of Ontario}}(2022{\natexlab{d}})}]{government_of_ontario_census_2022-3}%
  \BibitemOpen
  \bibfield  {author} {\bibinfo {author} {\bibnamefont {{Government of Ontario}}},\ }\href {http://www.ontario.ca/document/2016-census-highlights/fact-sheet-1-population-counts-canada-ontario-and-regions} {{\selectlanguage {en}\bibinfo {title} {Census {Fact} {Sheet} 1: {Population} counts: {Canada}, {Ontario} and regions {\textbar} 2016 census highlights}}} (\bibinfo {year} {2022}{\natexlab{d}})\BibitemShut {NoStop}%
\bibitem [{\citenamefont {{Statistics Canada}}(2017)}]{statistics_canada_postal_2017}%
  \BibitemOpen
  \bibfield  {author} {\bibinfo {author} {\bibnamefont {{Statistics Canada}}},\ }\href {https://www150.statcan.gc.ca/n1/pub/92-154-g/92-154-g2017001-eng.htm} {\bibinfo {title} {Postal {Code} {OM} {Conversion} {File} ({PCCF}), {Reference} {Guide}, 2017}} (\bibinfo {year} {2017}),\ \bibinfo {note} {last Modified: 2017-12-13}\BibitemShut {NoStop}%
\bibitem [{\citenamefont {of~Education}(2018)}]{ontario_ministry_of_education_school_2018}%
  \BibitemOpen
  \bibfield  {author} {\bibinfo {author} {\bibfnamefont {O.~M.}\ \bibnamefont {of~Education}},\ }\href {https://data.ontario.ca/dataset/school-information-and-student-demographics} {\bibinfo {title} {School information and student demographics - {Dataset} - {Ontario} {Data} {Catalogue}}} (\bibinfo {year} {2018})\BibitemShut {NoStop}%
\bibitem [{\citenamefont {Dooley}\ \emph {et~al.}(2017)\citenamefont {Dooley}, \citenamefont {Payne}, \citenamefont {Steffler},\ and\ \citenamefont {Wagner}}]{dooley_understanding_2017}%
  \BibitemOpen
  \bibfield  {author} {\bibinfo {author} {\bibfnamefont {M.}~\bibnamefont {Dooley}}, \bibinfo {author} {\bibfnamefont {A.}~\bibnamefont {Payne}}, \bibinfo {author} {\bibfnamefont {M.}~\bibnamefont {Steffler}},\ and\ \bibinfo {author} {\bibfnamefont {J.}~\bibnamefont {Wagner}},\ }\bibfield  {title} {\bibinfo {title} {Understanding the {STEM} {Path} through {High} {School} and into {University} {Programs}},\ }\href {http://muse.jhu.edu/article/651475} {\bibfield  {journal} {\bibinfo  {journal} {Canadian Public Policy}\ }\textbf {\bibinfo {volume} {43}},\ \bibinfo {pages} {1} (\bibinfo {year} {2017})},\ \bibinfo {note} {publisher: University of Toronto Press}\BibitemShut {NoStop}%
\bibitem [{Note1()}]{Note1}%
  \BibitemOpen
  \bibinfo {note} {The term Visible Minority was defined by the Employment Equity Act as “persons, other than Aboriginal peoples, who are non-Caucasian in race or non-white in colour”. Beyond this section where we use variable names as given by the census, we chose to use the more widely accepted terms Person of Colour or Racialized Individual instead of Visible Minority and Indigenous instead of Aboriginal.}\BibitemShut {Stop}%
\bibitem [{\citenamefont {for Health~Information}(2020)}]{canadian_institute_for_health_information_proposed_2020}%
  \BibitemOpen
  \bibfield  {author} {\bibinfo {author} {\bibfnamefont {C.~I.}\ \bibnamefont {for Health~Information}},\ }\href@noop {} {{\selectlanguage {en}\emph {\bibinfo {title} {Proposed {Standards} for {Race}-{Based} and {Indigenous} {Identity} {Data} {Collection} and {Health} {Reporting} in {Canada}}}}},\ \bibinfo {type} {Tech. Rep.}\ (\bibinfo  {institution} {Canadian Institute for Health Information},\ \bibinfo {address} {Ottawa, ON},\ \bibinfo {year} {2020})\BibitemShut {NoStop}%
\bibitem [{\citenamefont {Statistics}(2022)}]{aip_statistics_initial_2022}%
  \BibitemOpen
  \bibfield  {author} {\bibinfo {author} {\bibfnamefont {A.}~\bibnamefont {Statistics}},\ }\href {https://www.aip.org/statistics/resources/initial-employment-physics-bachelors-and-phds-classes-2019-and-2020} {{\selectlanguage {en}\emph {\bibinfo {title} {Initial employment: physics bachelors and phds classes of 2019 and 2020}}}},\ \bibinfo {type} {Tech. Rep.}\ (\bibinfo  {institution} {American Insitute of Physics},\ \bibinfo {year} {2022})\BibitemShut {NoStop}%
\bibitem [{\citenamefont {Turcotte}(2011)}]{turcotte_intergenerational_2011}%
  \BibitemOpen
  \bibfield  {author} {\bibinfo {author} {\bibfnamefont {M.}~\bibnamefont {Turcotte}},\ }\href {https://www150.statcan.gc.ca/n1/pub/11-008-x/2011002/article/11536-eng.htm} {{\selectlanguage {en}\emph {\bibinfo {title} {Intergenerational education mobility: {University} completion in relation to parents' education level}}}},\ \bibinfo {type} {Tech. Rep.}\ \bibinfo {number} {11-008-X}\ (\bibinfo  {institution} {Statistics Canada},\ \bibinfo {year} {2011})\BibitemShut {NoStop}%
\bibitem [{\citenamefont {{Statistics Canada}}(2016{\natexlab{a}})}]{statistics_canada_national_2016}%
  \BibitemOpen
  \bibfield  {author} {\bibinfo {author} {\bibnamefont {{Statistics Canada}}},\ }\href {https://noc.esdc.gc.ca/} {{\selectlanguage {eng}\bibinfo {title} {National {Occupational} {Classification} ({NOC})}}} (\bibinfo {year} {2016}{\natexlab{a}}),\ \bibinfo {note} {last Modified: 2021-11-30}\BibitemShut {NoStop}%
\bibitem [{\citenamefont {{Statistics Canada}}(2016{\natexlab{b}})}]{statistics_canada_classification_2016}%
  \BibitemOpen
  \bibfield  {author} {\bibinfo {author} {\bibnamefont {{Statistics Canada}}},\ }\href {https://www.statcan.gc.ca/en/subjects/standard/cip/2016/introduction} {{\selectlanguage {en}\bibinfo {title} {Classification of {Instructional} {Programs} ({CIP})}}} (\bibinfo {year} {2016}{\natexlab{b}})\BibitemShut {NoStop}%
\bibitem [{\citenamefont {Stine}(1995)}]{stine_graphical_1995}%
  \BibitemOpen
  \bibfield  {author} {\bibinfo {author} {\bibfnamefont {R.~A.}\ \bibnamefont {Stine}},\ }\bibfield  {title} {{\selectlanguage {English}\bibinfo {title} {Graphical interpretation of variance inflation factors}},\ }\href {https://www.proquest.com/docview/228425057/abstract/BB412CF5E59D4893PQ/1} {\bibfield  {journal} {\bibinfo  {journal} {The American Statistician}\ }\textbf {\bibinfo {volume} {49}},\ \bibinfo {pages} {53} (\bibinfo {year} {1995})},\ \bibinfo {note} {num Pages: 4 Place: Alexandria, United States Publisher: American Statistical Association}\BibitemShut {NoStop}%
\bibitem [{\citenamefont {{R Core Team}}(2023)}]{r_core_team_r_2023}%
  \BibitemOpen
  \bibfield  {author} {\bibinfo {author} {\bibnamefont {{R Core Team}}},\ }\href {https://www.r-project.org/} {\bibinfo {title} {R: {A} {Language} and {Environment} for {Statistical} {Computing}\vphantom{\{}\}}} (\bibinfo {year} {2023})\BibitemShut {NoStop}%
\bibitem [{\citenamefont {Wickham}\ \emph {et~al.}(2019)\citenamefont {Wickham}, \citenamefont {Averick}, \citenamefont {Bryan}, \citenamefont {Chang}, \citenamefont {McGowan}, \citenamefont {François}, \citenamefont {Grolemund}, \citenamefont {Hayes}, \citenamefont {Henry}, \citenamefont {Hester}, \citenamefont {Kuhn}, \citenamefont {Pedersen}, \citenamefont {Miller}, \citenamefont {Bache}, \citenamefont {Müller}, \citenamefont {Ooms}, \citenamefont {Robinson}, \citenamefont {Seidel}, \citenamefont {Spinu}, \citenamefont {Takahashi}, \citenamefont {Vaughan}, \citenamefont {Wilke}, \citenamefont {Woo},\ and\ \citenamefont {Yutani}}]{wickham_welcome_2019}%
  \BibitemOpen
  \bibfield  {author} {\bibinfo {author} {\bibfnamefont {H.}~\bibnamefont {Wickham}}, \bibinfo {author} {\bibfnamefont {M.}~\bibnamefont {Averick}}, \bibinfo {author} {\bibfnamefont {J.}~\bibnamefont {Bryan}}, \bibinfo {author} {\bibfnamefont {W.}~\bibnamefont {Chang}}, \bibinfo {author} {\bibfnamefont {L.~D.}\ \bibnamefont {McGowan}}, \bibinfo {author} {\bibfnamefont {R.}~\bibnamefont {François}}, \bibinfo {author} {\bibfnamefont {G.}~\bibnamefont {Grolemund}}, \bibinfo {author} {\bibfnamefont {A.}~\bibnamefont {Hayes}}, \bibinfo {author} {\bibfnamefont {L.}~\bibnamefont {Henry}}, \bibinfo {author} {\bibfnamefont {J.}~\bibnamefont {Hester}}, \bibinfo {author} {\bibfnamefont {M.}~\bibnamefont {Kuhn}}, \bibinfo {author} {\bibfnamefont {T.~L.}\ \bibnamefont {Pedersen}}, \bibinfo {author} {\bibfnamefont {E.}~\bibnamefont {Miller}}, \bibinfo {author} {\bibfnamefont {S.~M.}\ \bibnamefont {Bache}}, \bibinfo {author} {\bibfnamefont {K.}~\bibnamefont {Müller}}, \bibinfo {author} {\bibfnamefont {J.}~\bibnamefont
  {Ooms}}, \bibinfo {author} {\bibfnamefont {D.}~\bibnamefont {Robinson}}, \bibinfo {author} {\bibfnamefont {D.~P.}\ \bibnamefont {Seidel}}, \bibinfo {author} {\bibfnamefont {V.}~\bibnamefont {Spinu}}, \bibinfo {author} {\bibfnamefont {K.}~\bibnamefont {Takahashi}}, \bibinfo {author} {\bibfnamefont {D.}~\bibnamefont {Vaughan}}, \bibinfo {author} {\bibfnamefont {C.}~\bibnamefont {Wilke}}, \bibinfo {author} {\bibfnamefont {K.}~\bibnamefont {Woo}},\ and\ \bibinfo {author} {\bibfnamefont {H.}~\bibnamefont {Yutani}},\ }\bibfield  {title} {{\selectlanguage {en}\bibinfo {title} {Welcome to the {Tidyverse}}},\ }\href {https://doi.org/10.21105/joss.01686} {\bibfield  {journal} {\bibinfo  {journal} {Journal of Open Source Software}\ }\textbf {\bibinfo {volume} {4}},\ \bibinfo {pages} {1686} (\bibinfo {year} {2019})}\BibitemShut {NoStop}%
\bibitem [{\citenamefont {Bates}\ \emph {et~al.}(2015)\citenamefont {Bates}, \citenamefont {Mächler}, \citenamefont {Bolker},\ and\ \citenamefont {Walker}}]{bates_fitting_2015}%
  \BibitemOpen
  \bibfield  {author} {\bibinfo {author} {\bibfnamefont {D.}~\bibnamefont {Bates}}, \bibinfo {author} {\bibfnamefont {M.}~\bibnamefont {Mächler}}, \bibinfo {author} {\bibfnamefont {B.}~\bibnamefont {Bolker}},\ and\ \bibinfo {author} {\bibfnamefont {S.}~\bibnamefont {Walker}},\ }\bibfield  {title} {{\selectlanguage {en}\bibinfo {title} {Fitting {Linear} {Mixed}-{Effects} {Models} {Using} \textbf{lme4}}},\ }\bibfield  {journal} {\bibinfo  {journal} {Journal of Statistical Software}\ }\textbf {\bibinfo {volume} {67}},\ \href {https://doi.org/10.18637/jss.v067.i01} {10.18637/jss.v067.i01} (\bibinfo {year} {2015})\BibitemShut {NoStop}%
\bibitem [{\citenamefont {Wilke}(2020)}]{wilke_cowplot_2020}%
  \BibitemOpen
  \bibfield  {author} {\bibinfo {author} {\bibfnamefont {C.~O.}\ \bibnamefont {Wilke}},\ }\href {https://cran.r-project.org/web/packages/cowplot/index.html} {\bibinfo {title} {cowplot: {Streamlined} {Plot} {Theme} and {Plot} {Annotations} for 'ggplot2'}} (\bibinfo {year} {2020})\BibitemShut {NoStop}%
\bibitem [{\citenamefont {Hlavac}(2022)}]{hlavac_stargazer_2022}%
  \BibitemOpen
  \bibfield  {author} {\bibinfo {author} {\bibfnamefont {M.}~\bibnamefont {Hlavac}},\ }\href {https://cran.r-project.org/web/packages/stargazer/index.html} {\bibinfo {title} {stargazer: {Well}-{Formatted} {Regression} and {Summary} {Statistics} {Tables}}} (\bibinfo {year} {2022})\BibitemShut {NoStop}%
\bibitem [{\citenamefont {Bartoń}(2023)}]{barton_mumin_2023}%
  \BibitemOpen
  \bibfield  {author} {\bibinfo {author} {\bibfnamefont {K.}~\bibnamefont {Bartoń}},\ }\href {https://cran.r-project.org/web/packages/MuMIn/index.html} {\bibinfo {title} {{MuMIn}: {Multi}-{Model} {Inference}}} (\bibinfo {year} {2023})\BibitemShut {NoStop}%
\bibitem [{\citenamefont {Nakagawa}\ \emph {et~al.}(2017)\citenamefont {Nakagawa}, \citenamefont {Johnson},\ and\ \citenamefont {Schielzeth}}]{nakagawa_coefficient_2017}%
  \BibitemOpen
  \bibfield  {author} {\bibinfo {author} {\bibfnamefont {S.}~\bibnamefont {Nakagawa}}, \bibinfo {author} {\bibfnamefont {P.~C.~D.}\ \bibnamefont {Johnson}},\ and\ \bibinfo {author} {\bibfnamefont {H.}~\bibnamefont {Schielzeth}},\ }\bibfield  {title} {\bibinfo {title} {The coefficient of determination {R2} and intra-class correlation coefficient from generalized linear mixed-effects models revisited and expanded},\ }\href {https://doi.org/10.1098/rsif.2017.0213} {\bibfield  {journal} {\bibinfo  {journal} {Journal of The Royal Society Interface}\ }\textbf {\bibinfo {volume} {14}},\ \bibinfo {pages} {20170213} (\bibinfo {year} {2017})},\ \bibinfo {note} {publisher: Royal Society}\BibitemShut {NoStop}%
\bibitem [{\citenamefont {Cooper}\ and\ \citenamefont {Berry}(2020)}]{cooper_demographic_2020}%
  \BibitemOpen
  \bibfield  {author} {\bibinfo {author} {\bibfnamefont {G.}~\bibnamefont {Cooper}}\ and\ \bibinfo {author} {\bibfnamefont {A.}~\bibnamefont {Berry}},\ }\bibfield  {title} {\bibinfo {title} {Demographic predictors of senior secondary participation in biology, physics, chemistry and earth/space sciences: students’ access to cultural, social and science capital},\ }\href {https://doi.org/10.1080/09500693.2019.1708510} {\bibfield  {journal} {\bibinfo  {journal} {International Journal of Science Education}\ }\textbf {\bibinfo {volume} {42}},\ \bibinfo {pages} {151} (\bibinfo {year} {2020})},\ \bibinfo {note} {publisher: Routledge}\BibitemShut {NoStop}%
\bibitem [{\citenamefont {Jin}(2021)}]{jin_supporting_2021}%
  \BibitemOpen
  \bibfield  {author} {\bibinfo {author} {\bibfnamefont {Q.}~\bibnamefont {Jin}},\ }\bibfield  {title} {{\selectlanguage {en}\bibinfo {title} {Supporting {Indigenous} {Students} in {Science} and {STEM} {Education}: {A} {Systematic} {Review}}},\ }\href {https://doi.org/10.3390/educsci11090555} {\bibfield  {journal} {\bibinfo  {journal} {Education Sciences}\ }\textbf {\bibinfo {volume} {11}},\ \bibinfo {pages} {555} (\bibinfo {year} {2021})},\ \bibinfo {note} {number: 9 Publisher: Multidisciplinary Digital Publishing Institute}\BibitemShut {NoStop}%
\bibitem [{\citenamefont {{Statistics Canada}}(2023)}]{statistics_canada_profile_2023}%
  \BibitemOpen
  \bibfield  {author} {\bibinfo {author} {\bibnamefont {{Statistics Canada}}},\ }\href {https://www12.statcan.gc.ca/census-recensement/2021/dp-pd/ipp-ppa/details/page.cfm?Lang=E&DGUID=2021A000011124&GENDER=1&AGE=1&RESIDENCE=1&HP=0&HH=0&SearchText=Canada} {{\selectlanguage {eng}\bibinfo {title} {Profile table: {Canada} [{Country}], {Indigenous} {Population} {Profile}, 2021 {Census} of {Population}}}} (\bibinfo {year} {2023}),\ \bibinfo {note} {last Modified: 2023-06-21}\BibitemShut {NoStop}%
\bibitem [{\citenamefont {James}\ and\ \citenamefont {Turner}(2017)}]{james_towards_2017}%
  \BibitemOpen
  \bibfield  {author} {\bibinfo {author} {\bibfnamefont {C.~E.}\ \bibnamefont {James}}\ and\ \bibinfo {author} {\bibfnamefont {T.}~\bibnamefont {Turner}},\ }\href {https://edu.yorku.ca/files/2017/04/Towards-Race-Equity-in-Education-April-2017.pdf?x38104} {{\selectlanguage {English}\emph {\bibinfo {title} {Towards {Race} {Equity} {In} {Education}: {The} {Schooling} of {Black} {Students} in the {Greater} {Toronto} {Area}}}}},\ \bibinfo {type} {Tech. Rep.}\ (\bibinfo  {institution} {York University},\ \bibinfo {address} {Toronto, Ontario},\ \bibinfo {year} {2017})\BibitemShut {NoStop}%
\bibitem [{\citenamefont {Dickens}\ \emph {et~al.}(2020)\citenamefont {Dickens}, \citenamefont {Jones},\ and\ \citenamefont {Hall}}]{dickens_being_2020}%
  \BibitemOpen
  \bibfield  {author} {\bibinfo {author} {\bibfnamefont {D.}~\bibnamefont {Dickens}}, \bibinfo {author} {\bibfnamefont {M.}~\bibnamefont {Jones}},\ and\ \bibinfo {author} {\bibfnamefont {N.}~\bibnamefont {Hall}},\ }\bibfield  {title} {\bibinfo {title} {Being a {Token} {Black} {Female} {Faculty} {Member} in {Physics}: {Exploring} {Research} on {Gendered} {Racism}, {Identity} {Shifting} as a {Coping} {Strategy}, and {Inclusivity} in {Physics}},\ }\href {https://doi.org/10.1119/1.5145529} {\bibfield  {journal} {\bibinfo  {journal} {The Physics Teacher}\ }\textbf {\bibinfo {volume} {58}},\ \bibinfo {pages} {335} (\bibinfo {year} {2020})},\ \bibinfo {note} {publisher: American Association of Physics Teachers}\BibitemShut {NoStop}%
\bibitem [{\citenamefont {Hyater-Adams}\ \emph {et~al.}(2018)\citenamefont {Hyater-Adams}, \citenamefont {Fracchiolla}, \citenamefont {Finkelstein},\ and\ \citenamefont {Hinko}}]{hyater-adams_critical_2018}%
  \BibitemOpen
  \bibfield  {author} {\bibinfo {author} {\bibfnamefont {S.}~\bibnamefont {Hyater-Adams}}, \bibinfo {author} {\bibfnamefont {C.}~\bibnamefont {Fracchiolla}}, \bibinfo {author} {\bibfnamefont {N.}~\bibnamefont {Finkelstein}},\ and\ \bibinfo {author} {\bibfnamefont {K.}~\bibnamefont {Hinko}},\ }\bibfield  {title} {{\selectlanguage {en}\bibinfo {title} {Critical look at physics identity: {An} operationalized framework for examining race and physics identity}},\ }\href {https://doi.org/10.1103/PhysRevPhysEducRes.14.010132} {\bibfield  {journal} {\bibinfo  {journal} {Physical Review Physics Education Research}\ }\textbf {\bibinfo {volume} {14}},\ \bibinfo {pages} {010132} (\bibinfo {year} {2018})}\BibitemShut {NoStop}%
\bibitem [{\citenamefont {Hyater-Adams}\ \emph {et~al.}(2019)\citenamefont {Hyater-Adams}, \citenamefont {Fracchiolla}, \citenamefont {Williams}, \citenamefont {Finkelstein},\ and\ \citenamefont {Hinko}}]{hyater-adams_deconstructing_2019}%
  \BibitemOpen
  \bibfield  {author} {\bibinfo {author} {\bibfnamefont {S.}~\bibnamefont {Hyater-Adams}}, \bibinfo {author} {\bibfnamefont {C.}~\bibnamefont {Fracchiolla}}, \bibinfo {author} {\bibfnamefont {T.}~\bibnamefont {Williams}}, \bibinfo {author} {\bibfnamefont {N.}~\bibnamefont {Finkelstein}},\ and\ \bibinfo {author} {\bibfnamefont {K.}~\bibnamefont {Hinko}},\ }\bibfield  {title} {\bibinfo {title} {Deconstructing {Black} physics identity: {Linking} individual and social constructs using the critical physics identity framework},\ }\href {https://doi.org/10.1103/PhysRevPhysEducRes.15.020115} {\bibfield  {journal} {\bibinfo  {journal} {Physical Review Physics Education Research}\ }\textbf {\bibinfo {volume} {15}},\ \bibinfo {pages} {020115} (\bibinfo {year} {2019})},\ \bibinfo {note} {publisher: American Physical Society}\BibitemShut {NoStop}%
\bibitem [{\citenamefont {McGee}(2016)}]{mcgee_devalued_2016}%
  \BibitemOpen
  \bibfield  {author} {\bibinfo {author} {\bibfnamefont {E.~O.}\ \bibnamefont {McGee}},\ }\bibfield  {title} {{\selectlanguage {en}\bibinfo {title} {Devalued {Black} and {Latino} {Racial} {Identities}: {A} {By}-{Product} of {STEM} {College} {Culture}?}},\ }\href {https://doi.org/10.3102/0002831216676572} {\bibfield  {journal} {\bibinfo  {journal} {American Educational Research Journal}\ }\textbf {\bibinfo {volume} {53}},\ \bibinfo {pages} {1626} (\bibinfo {year} {2016})},\ \bibinfo {note} {publisher: American Educational Research Association}\BibitemShut {NoStop}%
\bibitem [{\citenamefont {Rosa}\ and\ \citenamefont {Mensah}(2016)}]{rosa_educational_2016}%
  \BibitemOpen
  \bibfield  {author} {\bibinfo {author} {\bibfnamefont {K.}~\bibnamefont {Rosa}}\ and\ \bibinfo {author} {\bibfnamefont {F.~M.}\ \bibnamefont {Mensah}},\ }\bibfield  {title} {\bibinfo {title} {Educational pathways of {Black} women physicists: {Stories} of experiencing and overcoming obstacles in life},\ }\href {https://doi.org/10.1103/PhysRevPhysEducRes.12.020113} {\bibfield  {journal} {\bibinfo  {journal} {Physical Review Physics Education Research}\ }\textbf {\bibinfo {volume} {12}},\ \bibinfo {pages} {020113} (\bibinfo {year} {2016})},\ \bibinfo {note} {publisher: American Physical Society}\BibitemShut {NoStop}%
\bibitem [{\citenamefont {George-Jackson}\ and\ \citenamefont {Lichtenberger}(2012)}]{george-jackson_college_2012}%
  \BibitemOpen
  \bibfield  {author} {\bibinfo {author} {\bibfnamefont {C.~E.}\ \bibnamefont {George-Jackson}}\ and\ \bibinfo {author} {\bibfnamefont {E.~J.}\ \bibnamefont {Lichtenberger}},\ }\href {https://eric.ed.gov/?id=ED544651} {{\selectlanguage {en}\emph {\bibinfo {title} {College {Confidence}: {How} {Sure} {High} {School} {Students} {Are} of {Their} {Future} {Majors}. {Policy} {Research}: {IERC} 2012-2}}}}\ (\bibinfo  {publisher} {Illinois Education Research Council},\ \bibinfo {year} {2012})\ \bibinfo {note} {iSSN: ISSN- Publication Title: Illinois Education Research Council}\BibitemShut {NoStop}%
\bibitem [{\citenamefont {Stoet}\ and\ \citenamefont {Geary}(2018)}]{stoet_gender-equality_2018}%
  \BibitemOpen
  \bibfield  {author} {\bibinfo {author} {\bibfnamefont {G.}~\bibnamefont {Stoet}}\ and\ \bibinfo {author} {\bibfnamefont {D.~C.}\ \bibnamefont {Geary}},\ }\bibfield  {title} {{\selectlanguage {en}\bibinfo {title} {The {Gender}-{Equality} {Paradox} in {Science}, {Technology}, {Engineering}, and {Mathematics} {Education}}},\ }\href {https://doi.org/10.1177/0956797617741719} {\bibfield  {journal} {\bibinfo  {journal} {Psychological Science}\ }\textbf {\bibinfo {volume} {29}},\ \bibinfo {pages} {581} (\bibinfo {year} {2018})},\ \bibinfo {note} {publisher: SAGE Publications Inc}\BibitemShut {NoStop}%
\end{thebibliography}%

\end{document}